\documentclass[useAMS,usenatbib]{mn2e}
\usepackage{graphicx}
\usepackage{txfonts}
\begin{document}

\title[Multi-frequency pulsar scatter time estimates ]
{The analysis of the largest sample of multi-frequency pulsar scatter time estimates.}
\author[Wojciech Lewandowski et al.]
{Wojciech Lewandowski,\thanks{E-mail: boe@astro.ia.uz.zgora.pl}
Magdalena Kowali{\'n}ska,
Jaroslaw Kijak\\
Institute of Astronomy, University of Zielona G\'ora , Szafrana~2, 65-246~Zielona~G\'ora, Poland}
\date{Accepted . Received ; in original form }
\maketitle

\begin{abstract}

We present our results of pulse broadening time estimates and the study of the frequency scaling of this quantity for 60 pulsars based on actual multi-frequency scattering estimates. This research was based on our own measurements, performed on the observational data and the profiles from various pulsar profile databases, as well as the scatter time measurements that were found in the literature. We were able to construct a database of over 60 pulsars with true multi-frequency $\alpha$ measurements, which allowed us to revise the previously proposed relations between the scatter time spectral slope and the dispersion measure (DM). We found that the deviations from theoretical predictions of the value of $\alpha$ appear for pulsars regardless of their DM, however the DM-averaged value of the scaling index is almost constant except for pulsars with very high DMs. Based on the obtained slopes we were also able to estimate the amount of scattering at the standard frequency of 1~GHz. We found that while the estimated standardized pulse broadening time increases with DM the relation seems to be much flatter than it was previously proposed, which suggests higher values of the scatter time for mid-DM pulsars, and lower values of expected pulse broadening for highly dispersed sources.

\end{abstract}

\label{firstpage}


\section{Introduction}
 \label{intro}

The temporal and angular broadening of the pulsar profiles is caused by the scattering of the radio waves emitted by a pulsar. This, along with other effects such as an interstellar dispersion of the signal, as well as the interstellar scintillations, happens in the ionized interstellar medium (ISM). Since the discovery of these phenomena various analyses, both observational and theoretical, tried to explain the properties of the interaction of the pulsar radiation with ISM. In our recent paper \citep[][hereafter Paper~1]{lewan13} we performed a statistical analysis of the scattering phenomenon, especially the scatter time frequency dependence, expanding upon an earlier work of \citet[][L01 and L04 hereafter]{L01,L04}.  Our analysis confirmed their findings that the frequency dependence of the scatter time measurements for the low dispersion measure (DM) pulsars seems to conform the theoretical predictions, while higher-DM pulsars apparently deviate from the theory, in such way that their spectral dependence is much flatter.

The scattering phenomenon theory \citep[see][for a review]{rickett90} predicts, that even if a pulse leaves the pulsar in a single instant, due to the propagation in the interstellar medium and the scattering (bending) of the rays the observer will measure this signal over a certain interval. Since the radiation that had undergone scattering was travelling along longer geometrical paths it arrives delayed, causing the apparent pulse to attain the so called ``scattering tail'' which is roughly consistent with an exponential decay of the signal. 

If one assumes the simplest geometrical model of the ISM, i.e. a configuration in which all the scattering happens within a thin screen of ionized medium, the shape of the tail may be predicted by assuming a two-dimensional gaussian brightness distribution of the scattering. In such case the resulting pulse broadening function ($PBF$) has a simple exponential decay form: $PBF(t) \sim \exp(t/\tau_d)$ where characteristic decay time $\tau_d$ is called the scatter time \citep[or pulse broadening time, ][]{scheuer68}. 

The brightness distribution of the scattering (and in return the shape of the scattering tail) can be estimated by assuming a model of the turbulent interstellar medium that is causing this phenomenon. Usually it is done by assuming the homogeneous isotropic turbulence \citep{rickett77} within a certain range of the fluctuation scales (the {\it inner scale} $k_i^{-1}$ and the {\it outer scale} $k_o^{-1}$). The spectrum of such turbulence can be described by:

\begin{equation}
\label{elec_dens_full}
P_{n_e} (q) = \frac{C_{n_e}^2}{(q^2 + k_o^2)\ ^{\beta/2}} \exp\left(-\frac{q^2}{4\ k_i^2}\right), 
\end{equation}

\noindent
where $q$ is an amplitude of a three-dimensional wave number and $C_{n_e}^2$ is the fluctuation strength for a given line-of-sight. If $k_o \ll q \ll k_i$ then the above equation simplifies to a power-law:

\begin{equation}
\label{elec_dens_simple}
P_{n_e} (q) = C_{n_e}^2 q^{-\beta},
\end{equation}

\noindent 

The fluctuation spectral index $\beta$ has to be lower than 4 \citep{romani86} and in such case the scatter time frequency dependence also takes a form of a power law: $\tau_d \propto \nu_{obs}^{-\alpha}$. The fluctuation spectral index $\beta$ and the scatter time frequency scaling index $\alpha$ are  bound by a relation: $\alpha = 2\beta/(\beta -2)$. One has to note that the scatter time frequency dependence slope is the same as the decorrelation bandwidth slope $\Delta\nu_d$ (apart for the minus sign), a quantity from the interstellar scintillation (ISS) theory, in which both these values are bound by a formula:

\begin{equation}
\label{def_c}
2\pi\, \tau_d\, \Delta\nu_d = C_1,
\end{equation}

\noindent
where the $C_1$ is a constant of the order of unity, although it may depend on the turbulence model or the ISM geometry \citep{lambert99}. 

If one adopts a purely Kolmogorov spectrum of the density irregularities ($\beta = 11/3$) the resulting value of the scatter time spectral index is $\alpha=4.4$. Other turbulence models may yield different values of $\alpha$, but the lowest value for a thin screen model is $\alpha=4.0$ (for the so called critical turbulence model with $\beta=4$, \citealt{romani86}).

As we pointed out in Paper~1 (following other authors) the simplification of Eq.~\ref{elec_dens_full} may not be valid, since we do not know what the inner/outer scales are. Especially the inner scale may be important, since when the diffractive scale $q^{-1}$ drops below $k_i^{-1}$ (and thus $q>k_i$) the dependence becomes quadratic. As the diffractive scale changes with the observing frequency, then  for a particular pulsar this may result in it having the theoretical scatter time slope at high frequencies, and the slope will become flatter at lower frequencies (i.e for Kolmogorov turbulence $\alpha$ would fall from 4.4 to 4.0). So far  the sparsity of multi-frequency scatter time measurements made it virtually impossible to measure such effects. There were however attempts to detect the inner scale influence by the means of the scattering tail shape analysis \citep[see for example ][]{rickett09}.

One has to also note that the measured slopes of the scatter time frequency dependence may be affected by other effects, the most important being the anisotropy and inhomogeneity of the ISM. The anisotropy may falsify single scatter measurement, since it may affect the shape of the scattering tail \citep{rickett09} and falsify the results of the shape modelling (yielding erroneous values of $\tau_d$). Inhomogeneity on the other hand can cause the scattering parameters to vary in time \citep[see ][and also our Paper~1]{brisken10}, which in turn can make any attempts of estimating the  scatter time spectrum impossible, if the observations at different frequencies were made at vastly different epochs.

The multi-frequency scatter time analysis  (often coupled with the decorrelation bandwidth estimates from interstellar scintillation observations) was attempted many times in the past. The most  significant examples are the works of \citet{cordes85}, who analysed the data for 76 pulsars, although for only 5 of them the authors were able to obtain proper estimates of the decorrelation bandwidth frequency scaling slope based on estimates from 3 or more frequencies. Later \citet{bhat04} reported  the analysis of the observational data for 98 pulsars, although most of them were measured at a single frequency and they were able to estimate the frequency scaling slopes only for 15 sources (based on data from 2 frequencies), however one has to mention that in the same paper they were able to gather and analyse a large database of 371 decorrelation bandwidth and pulse broadening estimates that was available at the time. Simultaneously L01 and L04 reported the scatter time frequency scaling analysis estimates based exclusively on pulsars which had the scatter time or decorrelation bandwidth estimates at three or more observing frequencies, and were able to gather 27 such pulsars - both from their own observations as well as from the literature. In our Paper~1 we decided to expand their work, adding a few more sources but not all of them had measurements on three frequencies. Hence  we expanded our database  by looking for any scatter time measurements available in the literature. 

In our research, following L01 and L04,  we focused on pulsars which had reliable pulse broadening time estimates on at least three observing frequencies, which allows for a proper frequency scaling analysis. To increase the number of such objects we revisited the available profile databases - the EPN database\footnote{\tt http://www.jb.man.ac.uk/pulsar/Resources/epn/} and the profiles obtained by Parkes telescope which can be found in the ATNF Database\footnote{The ANDSATNF database, which is maintained and distributed by the means of CSIRO Research Data Service is available at \tt https://data.csiro.au/dap/search?tn=Astronomical and Space Sciences not elsewhere classified} - and measured scatter times for selected pulsars. Finally we supplemented all the above by adding the results from our newest GMRT and Effelsberg observing projects (which involved searching for gigahertz-peaked spectra pulsars, see \citealt{kijak11}). This yielded a number of about 100 pulsars with multi-frequency scatter time measurements, of which about 40 sources (including the ones already published in Paper~1) had reliable measurements at three or more frequencies. In this paper we present the results of our study. Adding our results to the 27 pulsars from L04 we were aiming to revise the L04 study of  the scatter time scaling index versus the dispersion measure for 60 objects in total, which to our knowledge. is the largest sample based on true multi-frequency estimates. These estimates allowed us also to extrapolate/interpolate the scattering strength to the standardized frequency of 1 GHz (similarly to the study  of \citealt{bhat04}) and we analysed its DM-dependence.

\section{Observations, Data Gathering and Analysis}
\label{obs}
To extend our the scatter time measurement database published in Paper~1 we searched all available literature, as well as added our own new measurements, both based on available pulsar profile databases, as well as from our own observations performed with the Giant Meterwave Radio Telescope (GMRT, Pune, India) at the frequencies of 325~MHz, 610~MHz and 1170~MHz.

The new GMRT observations were conducted between 2010 and 2012 as a part of our ongoing project to search for the gigahertz peaked spectra pulsars \citep{kijak11,dembska14}. We used 16~MHz bandwidth of GMRT working in phased array mode,  0.512 ms sampling rate and 256 spectral channels \citep[see ][]{guptaetal00}. The main goal of this observing project was to measure the flux densities of selected sources (to estimate the shape of their spectra), hence  
the data was intensity-calibrated using measurements of known continuous sources (like 1822-096). Pulsar profiles were obtained by folding of 20-30 minute integrations (up to 1 hour in some cases). We estimate that the sensitivity of these observations were of the order of 1~mJy at the frequency of 325 MHz up to 0.5~mJy at both 610~MHz and 1060~MHz.
Such sensitivity was usually more than enough to measure the pulsars flux densities, as well as it allowed for reliable study of the pulse profiles.

Another source of pulse profile shapes to analyse was the archived profiles that can be found in two databases in the internet, namely the EPN Database and the Parkes ATNF Database.  While the latter we used mainly for the L-band 
profiles (1.3 to 1.6 GHz), the EPN database contains data from a very wide range of frequencies (at least for some of the sources), however some of the very old data were not usable due to poor signal-to-noise ratio and/or poor time resolution of the profiles. In our analysis we omitted such profiles since the resulting values of the scatter times may have been unreliable, hence bringing more of a confusion than a valuable extension of the database.

Due to the nature of the available profile data, both from our own observations and for the profiles obtained from databases, we could not  ensure the data from various observing frequencies to be (quasi-)simultaneous, which would be a proper way to construct the pulse broadening spectra  (as we suggested in Paper~1). Especially for the profiles taken from the databases the differences between the epochs of the observation at different frequencies may be of the order of several years. This makes the data prone to the inhomogeneity effects. This needs to be taken into account  when interpreting the results, and it is even more of a concern for our next source of data.

We have also collected the scatter time measurements from several publications - see the $\tau_d$ in Table~\ref{tau_sc_table} in the  appendix for a full reference list. The specifics concerning the methods used for obtaining the scatter time measurements can be found in their respective references. In some cases (which was especially true for the older publications) the authors did not always quote the measurement uncertainties. To counter that we  assumed  the uncertainties to be at the level of 30\%, which is probably an overestimation in most cases, but we found it to be appropriate at least for some of the lowest frequency measurements. Another problem with some of the  measurements is that the authors of some of the oldest publications we used did not publish the profiles upon which their measurements were based, so we could not asses their quality. This has to be taken into account, and as a result we were forced to disregard some of the measurements from our scatter time frequency dependence  models.

\subsection{Data Analysis and Results}

For the profiles we had available (both from our observations as well as from the EPN and Parkes databases) we performed the measurements of the scatter time using the same method as in Paper~1, which is basically following the method used by L01 and L04.  We have to note that there are other methods which have been shown successful for finding the scatter time for pulsars. One of them is the ``CLEAN'' method developed by \citet{bhat03} which does not require any assumptions on the intrinsic pulse shape and shows very promissing results \citep[see ][]{bhat04}. One has to mention also the cyclic spectral analysis method proposed by \citet{demorest11} , this method however requires access to the raw voltage data we did not have. We decided however against these methods, since more than often we were forced to combine our results with older $\tau_d$ measurements taken from the literature (to estimate the slope of the scatter time frequency dependence). We decided that it will be more reliable if we used the same (or at least similar) methods in our scatter time estimates. Additionally L01 demonstrated that at least in case of higher-DM pulsars the results from their method and the ``CLEAN'' method are usually consistent within uncertainties.

The apparent pulse profile is a convolution of the intrinsic profile shape $P^I(t)$ with functions representing several effects that affect the radiation, such as the scattering smearing function in the ISM $s(t)$, the ISM dispersion smearing function $d(t)$ and the receiving backend response function $i(t)$.
Thus, the observed profile shape $P^O$ may be obtained as \citep{rama97}:

\begin{equation}
\label{profile_shape}
P^O(t) = P^I(t) *  s(t) *  d(t) * i(t),
\end{equation}

\noindent
where the asterisks ($*$) denote the convolution.

The last two factors in this equation may become insignificant. The modern receiver systems usually are fast enough for the effect of $i(t)$ to be negligible. For the coherent dedispersion receivers, or for filterbank receivers with large number of spectral channels the dispersion smearing function should also not pose a problem, especially at higher frequencies. The observations at lower frequencies may suffer the effect, and hence the low-frequency scatter measurements may be falsified, especially if the actual scattering smearing is similar to the dispersion smearing within a single spectral channel. The dispersion smearing may also be a factor at higher observing frequencies, when using filterbanks with only a few spectral channels (as it was the case for our measurements with Effelsberg Radio Telescope; see the discussion in Paper~1), 
and it may also play a role for some of the older archived profile data.

The term in the convolution (Eq.~\ref{profile_shape}) that solely interest us is the $s(t)$ function, which describes the broadening of a pulse due to the interstellar scattering. In the past various authors used various mathematical forms of this function. In our analysis we followed the simplest approach  which was shown by  L04, i.e. that a simple exponential decay function seems to work the best, at least for higher DM pulsars. This function describes scattering in the single thin screen model  \citep[see][]{williamson72, williamson73}, and can be expressed as (L04):

\begin{equation}
\label{pbf1}
PBF_1(t) = \exp(-t/\tau_d) U(t),
\end{equation}

\noindent
where $U(t)$ is the unit step function ($U(t<0)=0$, and $U(t \geq 0) = 1$).
 
 \begin{figure*}
\resizebox{\hsize}{!}{\includegraphics{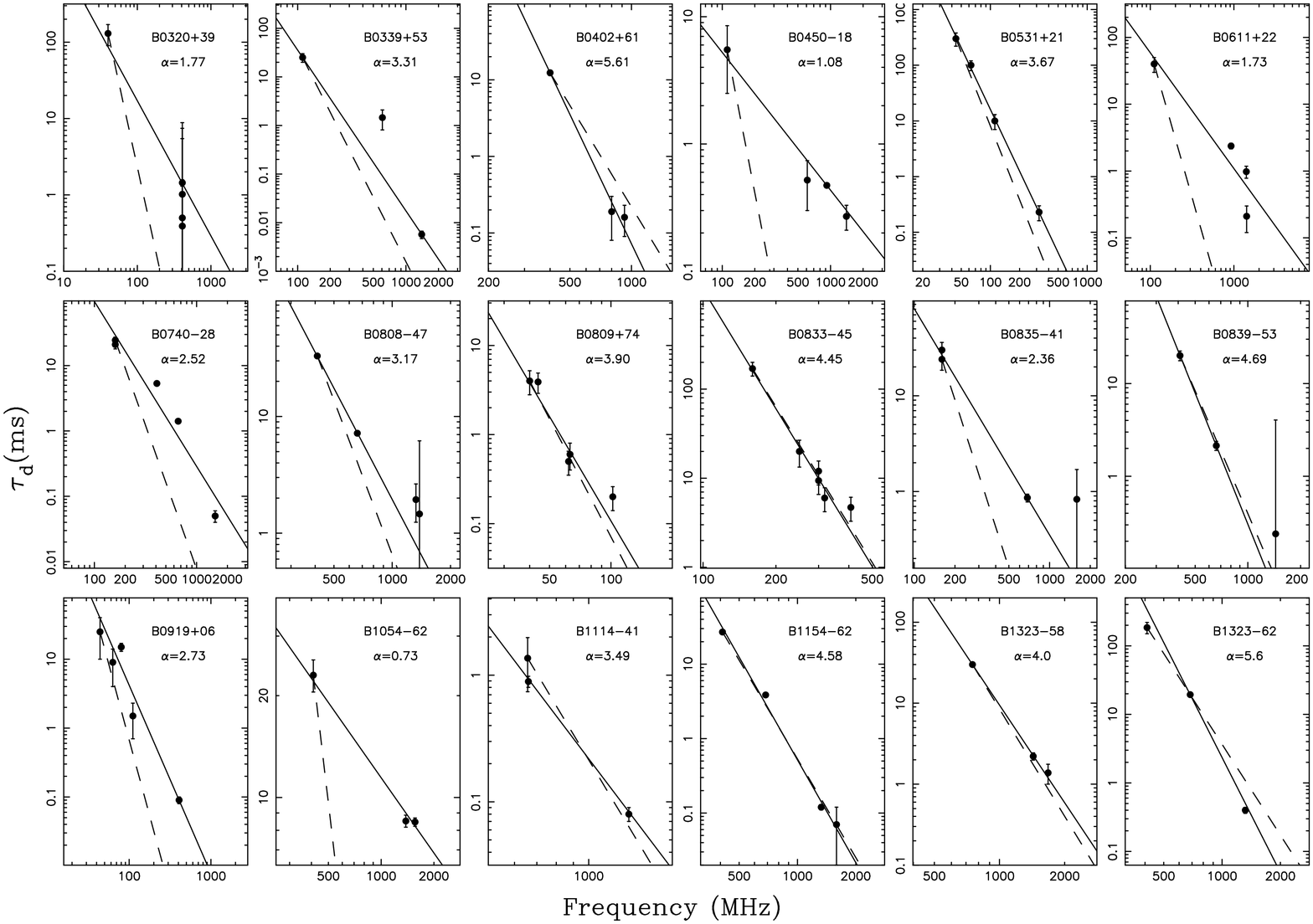}}
\caption{A plot of $\tau_{d}$ versus the observing frequency for pulsars from our sample. 
The solid line on each plot is the representation of the power-law fit to the scatter time measurements, while the dashed line shows the slope of the thin-screen Kolmogorov prediction with $\alpha = 4.4$.  ({\it Figure continued on the next pages}). 
\label{plot_tau_vs_freq}}
\end{figure*}

To find the values of $\tau_d$ for the observed profiles we performed a least squares fit of a model shape to the observational data, to find a best fit parameters by minimizing the $\chi^2$ function defined by the differences between the observed and modelled profiles, The fit was performed using the {\it Origin} software\footnote{http://www.originlab.com/} assuming a gaussian shape of the intrinsic pulse profile, or multiple gaussians in case of multi-component profiles. As it was discussed in Paper~1 such approach may be viewed as simplistic, but we do not know the intrinsic pulse profiles. This is a problem even if one uses the non-scattered high frequency profiles as templates. Such approach may suffer from various sources of error, such as the unknown profile evolution (which is cloaked by the scattering smear) and the radius-to-frequency mapping \citep{kijak03}\footnote{In many cases to satisfy the quality of the fit we had to allow for the width of the gaussian template to change in the fitting procedure, especially at lower frequencies. This supposedly should take care of the pulse width variations occurring due to radius-to-frequency mapping}. In Paper~1 we thoroughly discussed the possible sources of $\tau_d$ estimation errors when using our method. The most obvious of them is certain intrinsic profile shapes, especially a case of asymmetric profile with an intrinsic scatter-like tail.
This may not affect the resulting scatter time measurements by much when the interstellar scattering is large (i.e. much greater than the asymmetry of the profile), however when moving to higher observing frequencies the measured value of $\tau_d$ will saturate at some lowest level (which will be the characteristic time scale of the profile's asymmetry) when the ``real'' scattering becomes negligible.

One has to also remember about the very large scattering cases, i.e. instances where the scatter time is a significant fraction of the pulsar rotational period, or even greater. As we discussed in Paper~1 in such cases the background level of the profile may be falsified which may lead to underestimation of the measured scatter time. There is however a way to salvage at least some of the information as we did in case of PSR~B1557$-$50 observed at 325~MHz (and for two other pulsars), by correcting the profile background level. To address this we carefully examined the profiles we were analysing, but in our current sample we did not find observations which required us to use this method. However, we suspect, that at least some of the published $\tau_d$ measurements we used may have suffered from this, especially at the lowest observing frequencies. This can be recognized in cases  where the quoted values of the scatter time do not seem to correspond with the values measured at higher frequencies. Such measurements were usually omitted in further analysis.

The full listing of the new data in our scatter time measurement database,  including our new measurements can be found in Table~\ref{tau_sc_table} (see also the online material). This data, along with the measurements published in Paper~1, as well as the data from L01, L04 and earlier experiments (see Paper~1) constitutes the most complete multi-frequency $\tau_d$ measurement database compiled so far, and this was the object of our further study.

\section{The frequency evolution of pulse broadening}
\label{section_evolution}

\begin{figure*}
\resizebox{\hsize}{!}{\includegraphics{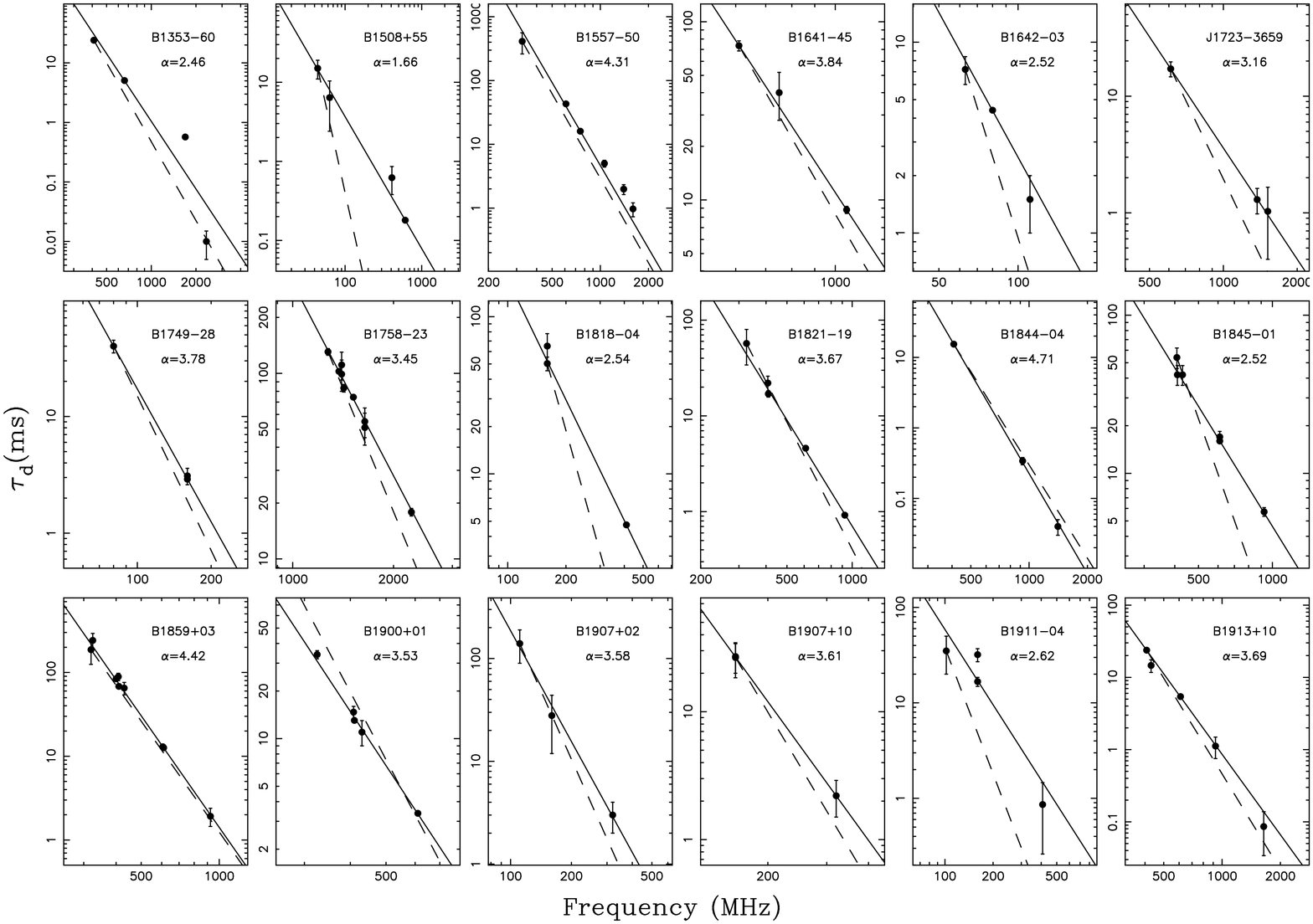}}
\vskip1mm
\centering{{\bf Figure~\ref{plot_tau_vs_freq}.} A plot of $\tau_{d}$ vs the observing frequency for pulsars from our sample (continued). 
\label{tau2}}
\end{figure*}

The main goal of our analysis was to construct the scatter pulse broadening spectra for as many pulsars as possible, to ascertain the frequency dependence of this quantity and compare it to the theoretical predictions. Using our database, which includes all the published data we could find as well as our new measurements (its acquisition was described in the previous section) we attempted to construct the scatter time spectra for about 80 pulsars. For the majority of them it was the first attempt to do so, for several other sources our current analysis added some individual $\tau_d$ measurements and forced us to re-evaluate the previously published spectra - see Paper~1 (and references therein) for a summary of scattering measurements prior to the results of our current analysis.

As we mentioned in the Introduction one can expect the measured scatter time values to be highly dependant on the observing frequency, and usually following a simple power-law. If one uses a simple thin screen model (or for that mater also for an uniformly distributed medium) with Kolmogorov type turbulence, then (excluding anisotropy, inhomogeneity and inner scale effects) one can expect the scatter time spectral index to be $\alpha = 4.4$. For other ISM geometries, or when the inner scale effects start to be significant one can expect the spectral index to drop down to $\alpha = 4.0$ or lower, even with the assumption of Kolmogorov fluctuation distribution in the ISM \citep[see for example ][]{cordes01}. 

To find the spectral slopes of the pulse broadening frequency dependence we modelled our data using a power-law. For the proper uncertainty analysis we performed weighted least-squares fits of the full power-law function:

 \begin{equation}
 \label{fit_form}
 \tau_d(\nu)= 10^{-\alpha \log \nu +b},
\end{equation} 
 \noindent
   instead of fitting a linear relation to the logarithms of $\tau_d$ and $\nu$. To perform weighted fits in cases where $\tau_d$ values did not have error estimates - which was often the case for some of the archival scatter time data, as well as for some more problematic of our pulse profile models - we assumed a 30\% uncertainty. 

Figure~\ref{plot_tau_vs_freq} shows the results of our modelling for 48 pulsars of the total number of about 80 pulsars we analysed in our current project. The remaining objects have either been measured at two frequencies only, or - due to the exclusion of some of the measured $\tau_d$ values  - were limited to either two data points or points at only two observing frequencies. As we mentioned in the Introduction, following L01 and L04 we decided to omit such results in our further analysis (beyond quoting the value of $\alpha$ in the table) as we were unable to reliably estimate the uncertainties of the spectral index values obtained from the fits.

We were also forced to omit some of the individual $\tau_d$ measurements from our fits as we found them unreliable. These were mainly the very old lowest frequency measurements (below 100 MHz) that did not conform higher frequency data, and in most cases were quoted without uncertainties in their original papers. We also disregarded some of the high frequency measurements (including some of our own) where we expected that the estimate was falsified by the limitations of our method - see the previous section and more detailed discussion in Paper~1. Using our method for profiles which show very little scattering yields the upper limits for the scatter time rather than actual estimates, and there is no reliable way to include these upper limits in our fits.

\begin{figure*}
\resizebox{\hsize}{!}{\includegraphics{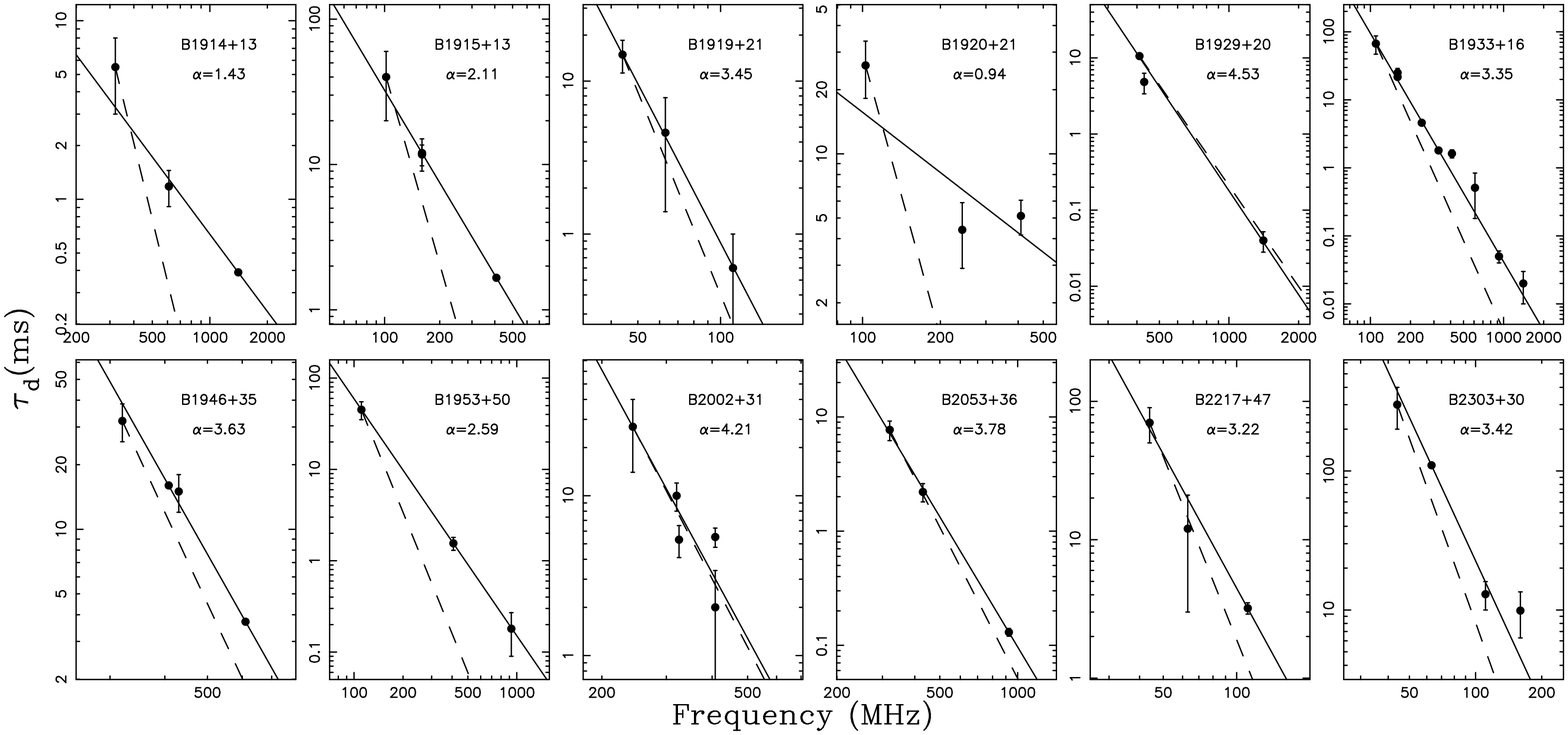}}
\vskip1mm
\centering{{\bf Figure~\ref{plot_tau_vs_freq}.} A plot of $\tau_{d}$ versus the observing frequency for pulsars from our sample (continued).\label{tau3}}
\end{figure*}


Table~\ref{table_alpha} summarizes the results of our fits for the pulsars in our sample, including only the sources for which the scattering was measured on at least three frequencies (full list of the scatter time measurements for all the pulsars from our database can be found in the online appendix). In this table, apart from the basic relevant pulsar data (that were taken from the ATNF Pulsar Catalogue\footnote{Available at {\tt http://www.atnf.csiro.au/people/pulsar/psrcat/}}, \citealt{manchester05}) we quote the the values of the pulse broadening time spectral index $\alpha$,  and the inferred turbulence spectral index $\beta$. The $\beta$ values were calculated using the $\beta=2 \alpha/(\alpha-2)$ relation, which is valid only for $\alpha \ge 4$, hence we do not present the resulting beta values when this relation is irrelevant (i.e. $\alpha$ within its uncertainties does not reach the value of 4). This does not mean however, that the values of $\alpha$ that are lower than 4.0 should be considered wrong or unreliable, only that within the constraints of the basic scattering model there is no way to estimate $\beta$ if $\alpha<4.0$, but such values are not ``disallowed'' in general (they are expected to appear for example for scattering geometries different from the thin screen model).

\begin{table*}
\caption{The scattering spectral index ($\alpha$) and the electron density spectral index ($\beta$) for the observed pulsars. Values of $\alpha$ quoted in italic are considered doubtful (see article text for explanation). Values of $\beta$ indicated by an asterisk are ``un-physical'', as the $\beta=2\alpha/(\alpha-2)$ relation is valid only for $\alpha \geq 4$. Table also lists the scattering fluctuation strength  $\log C_{n_e}^2 $ and the estimated scattering at a standard frequency of 1~GHz. In case of both of these the values given in brackets are considered unreliable (calculated based upon a doubtful value of $\alpha$). \label{table_alpha}}
\begin{tabular}{lrrrrllcc}\hline
Pulsar & $l_{II}~~$ &  $b_{II}~~$ & $DM$~~~~~ & Distance & $\alpha$~~ & $\beta$~~ & $\log C_{n_e}^2 $ & $\tau_d$ (ms)\\
      & (deg) & (deg) & (pc cm$^{-3}$) & (kpc)~~~ & & & &at 1 GHz\\ \hline \hline
    B0320+39     &    152.18  &  -14.34   &      26.764   &   1.49  &    {\it 1.77$\pm$ 0.2}   &    *     & {\it($-$1.36)}&{\it (0.29)}\\
    B0339+53     &    147.02  &   -1.43   &      67.30   &   2.48  &    3.31$\pm$0.2   &     *      & $-$3.01&0.0174\\
    B0402+61     &    144.02  &    7.05   &      65.303  &   3.05  &    5.61$\pm$0.5   &     3.1$\pm$ 0.5      & $-$2.86&0.0725\\
    B0450$-$18     &    217.08   &  -34.09   &      39.903  &   0.4   &     {\it1.08$\pm$0.2}    &   *      & {\it($-$0.43)}& {\it (0.434)}\\
    B0531+21     &    184.56  &   -5.78   &      56.791  &   2.00  &    4.21$\pm$0.5  &     3.8$\pm$0.9      & $-$4.12&0.00071\\
    B0611+22     &    188.79  &    2.40   &      96.91   &   4.74  &     {\it1.73$\pm$0.5}   &  *    & {\it($-$2.07)}&{\it (1.08)}\\
   B0740$-$28      &   243.77   &  -2.44    &     73.782   &  2.00   &   {\it 2.52$\pm$0.3}    &   *      & {\it($-$1.86)}&{\it (0.28)}\\
    B0808$-$47     &    263.30  &   -7.96   &     228.3    &  12.71  &    3.17$\pm$0.16  &     *     & $-$1.43&1.94\\
    B0809+74     &    140.00  &   31.62   &       6.116  &   0.43  &    3.90$\pm$0.85 &   4.1$\pm$1.8     & $-$4.76& 0.00025\\
    B0833$-$45     &    263.55  &   -2.79   &      67.99   &   0.28  &    4.45$\pm$0.35   &    3.63$\pm$0.57      & $-$0.98&0.047\\
    B0835$-$41     &    260.90  &   -0.34   &     147.29   &   1.50  &     {\it2.36$\pm$0.09}   &   *     & {\it($-$1.56)}&{\it (0.353)}\\
    B0839$-$53     &    270.77  &   -7.14   &     156.5    &   7.77  &    4.69$\pm$0.02  &     3.5$\pm$0.02      & $-$3.00&0.304\\
    B0919+06     &    225.42  &   36.39   &      27.271  &   1.10  &     {\it2.73$\pm$0.3}   &    *     & {\it($-$2.70)}& {\it (0.0077)}\\
    B1054$-$62     &    290.29  &   -2.97   &     320.3    &   2.40  &     {\it0.73$\pm$0.15}  &   *    & {\it($-$0.68)}&{\it (11.4)}\\
   B1114$-$41      &   284.45   &  18.07    &     40.53    &  2.68   &   3.50$\pm$0.17   &   *       & $-$1.66&0.210\\
   B1154$-$62      &   296.71   &  -0.20    &    325.2     &  4.00   &   4.59$\pm$0.27   &    3.54$\pm$0.42      &$-$2.30&0.493 \\
   B1323$-$58      &   307.50   &   3.56    &    287.30    &  3.00   &   4.00$\pm$0.08   &    4.00$\pm$0.16      & $-$0.48&9.47\\
   B1323$-$62      &   307.07   &   0.20    &    318.80    &  4.00   &   5.57$\pm$0.55   &    3.12$\pm$0 .62       & $-$2.19&2.35\\
   B1356$-$60      &   311.24   &   1.13    &    293.71    &  5.00   &   3.77$\pm$0.39   &    4.26$\pm$0.88      & $-$1.68&1.04\\
    B1508+55     &     91.33  &   52.29   &      19.613  &   2.10  &     {\it1.66$\pm$0.08}   &  *     & {\it($-$2.37)}&{\it (0.079)}\\
   B1557$-$50      &   330.69   &   1.63    &    260.56    &  6.90   &   4.31$\pm$0.36   &    3.73$\pm$0.62      &$-$1.78&4.64 \\
   B1641$-$45      &   339.19   &  -0.19    &    478.8     &  4.50   &   3.84$\pm$0.08   &    4.17$\pm$0.18    &$-$0.54&11.0 \\
    B1642$-$03     &     14.11  &   26.06   &      35.727  &   2.91  &    {\it 2.52$\pm$0.54}  &    *      &  {\it($-$3.48)}&{\it (0.0076)}\\
   J1723$-$3659    &   350.68   &  -0.41    &    254.2     &  4.28   &   3.16$\pm$0.04   &    *     & 0.32 &3.59\\
   B1749$-$28      &     1.54   &  -0.96    &     50.372   &  0.20   &   3.78$\pm$0.08   &    4.24 $\pm$0.18      &$-$1.60 &0.0028\\
      B1758$-$23      &     6.84   &  -0.07    &   1073.9     &  4.00   &   3.45$\pm$0.09   &    *      & 2.33& 306.9\\
    B1818$-$04     &     25.46  &    4.73   &      84.435  &   0.30  &     {\it2.54$\pm$0.10}  &    *    & {\it($-$0.16)}& {\it (0.488)}\\
   B1821$-$19      &    12.28   &  -3.11    &    224.648   &  3.70   &   3.67$\pm$0.11   &    *     & {\it($-$1.49)} &0.707\\            
   B1844$-$04      &    28.88   &  -0.94    &    141.979   &  3.12   &   4.71$\pm$0.08   &    3.48$\pm$0.12      &$-$1.80 & 0.224\\
    B1845$-$01     &     31.34  &    0.04   &     159.531  &   4.40  &    {\it 2.52$\pm$0.09}  &  *     & {\it($-$1.49)}& {\it (4.61)}\\
   B1859+03      &    37.21   &  -0.64    &    402.080   &  7.00   &   4.42$\pm$0.07   &    3.65$\pm$0.11      &$-$2.30&1.41 \\
   B1900+01      &    35.73   &  -1.96    &    245.167   &  3.30   &   3.53$\pm$0.10   &   *    &$-$1.32 & 0.58\\
   B1907+02      &    37.60   &  -2.71    &    171.734   &  4.50   &   3.58$\pm$0.24   &    *     & $-$3.05& 0.049\\
   B1907+10      &    44.83   &   0.99    &    149.982   &  4.80   &   3.61$\pm$0.03   &    *    &$-3.28$&0.036 \\
   B1911$-$04      &    31.31   &  -7.12    &     89.385   &  3.22   &    2.62$\pm$0.86   &   *     & $-$2.50&0.141\\
   B1913+10      &    44.71   &  -0.65    &    241.693   &  7.00   &   3.69$\pm$0.12   &    *     & $-$2.02&0.861\\
    B1914+13     &     47.58  &    0.45   &     237.009  &   4.50  &     {\it1.43$\pm$0.22}  &   *    & {\it($-$2.22)}& {\it (0.637)}\\
    B1915+13     &     48.26  &    0.62   &      94.538  &   5.00  &    {\it 2.11$\pm$0.05 } &     *     & {\it($-$2.64)}&{\it (0.249)}\\    
       B1919+21      &    55.78   &   3.50    &     12.455   &  0.30   &   3.45$\pm$0.07   &    *      & $-$3.26&0.00031\\    B1920+21     &     55.28  &    2.93   &     217.086  &   4.00  &    {\it 0.94$\pm$0.88} &   *      & {\it($-$1.19)}&{\it (7.46)}\\
     B1929+20      &    55.57   &   0.64    &    211.151   &  5.00   &   4.53$\pm$0.65   &    3.58$\pm$1.0       &$-$2.80 &0.179\\
   B1933+16      &    52.44   &  -2.09    &    158.521   &  3.70   &   3.35$\pm$0.15   &    *     & $-$2.89&0.0417\\
   B1946+35      &    70.70   &   5.05    &    129.075   &  7.87   &   3.63$\pm$0.06   &  *     & $-$2.24&0.619\\
    B1953+50     &     84.79  &   11.55   &      31.974  &   1.80  &     {\it2.59$\pm$0.01}  &    *     &  {\it($-$2.01)}&{\it (0.151)}\\
   B2002+31      &    69.01   &   0.02    &    234.820   &  8.00   &   4.21$\pm$0.66   &    3.81$\pm$1.19      &$-3.45$ &0.0689\\
   B2053+36      &    79.13   &  -5.59    &     97.3140  &  5.00   &   3.78$\pm$0.08   &    *      & $-2.85$&0.0964\\
   B2217+47      &    98.38   &  -7.60    &     43.519   &  2.45   &   3.22$\pm$0.32   &    *     & $-4.33$&2.67\\
   B2303+30      &    97.72   & -26.66    &     49.544   &  3.92   &   3.42$\pm$0.26   &    *       & $-3.94$&8.50\\
  \hline
\end{tabular}
\end{table*}

As one can see from the table the obtained values of $\alpha$ range from 0.73$\pm$0.15 for PSR~B1054$-$62 to 5.61$\pm$0.50 for PSR~B0402+61. Especially the very low values of $\alpha$ may be a source of concern and we examined all such cases closely. The most common reason behind such low values were the complicated morphologies of the profiles (in most cases clearly multi-component or asymmetric ones)  which may have influenced our fits to the data. One such case was the mentioned B1920+21 (see also Fig.~\ref{plot_tau_vs_freq} with an asymmetric profile and actually no significant scattering visible in any of the analysed profiles. In such cases one could say that our method measured not the evolution of scattering with observing frequency, but the frequency evolution of the intrinsic profile. Similar was the reason for excluding B1845$-$01. Another variation of this are cases where we had reliable measurements at lower observing frequencies but only other measurements were at very high frequencies where scattering was insignificant, but our method still gave us some values due to profile asymmetry. One such case is definitely PSR~B1054$-$62 (see Fig.~\ref{plot_tau_vs_freq} and pulse broadening measurements in the online appendix) for which we had data at 410~MHz with a reliable measurement, but the only 2 other data points are from our measurements performed for the profiles taken from the EPN Database at 1372~MHz and 1590~MHz. These profiles appear single but highly asymmetric and apparently the asymmetry is intrinsic and not related to scattering.

As we mentioned above, in our further scatter time frequency dependence analysis we decided to exclude all the pulsars which had scattering measured at only two frequencies and additionally for most of them the values of $\alpha$ are suspiciously low.
We also excluded also several other pulsars with low values of $\alpha$, like the pulsars which were measured at three frequencies but two of them were  relatively close, or pulsars with multi-component profiles. All the $\alpha$ values that we omitted are quoted in italic in Table~\ref{table_alpha}. We believe that the remaining 33 estimates of the scatter time frequency scaling index $\alpha$ are the most reliable estimates obtained for these pulsars so far.

Our final sample of pulsars with useful scatter time spectra estimations includes three pulsars that we already analysed in Paper~1, and that is because this time  new profiles for these pulsars and/or obtained values of $\tau$ from the literature that we did not use previously. The latter was true in case of PSR~B1557$-$50, where we added a measurement from \citet{rickett77} at 750 MHz. The addition of that measurement changed the $\alpha$ from 3.81$\pm$0.09 that was quoted in Paper~1 to 4.31$\pm$0.36 in our current analysis. This shows how sensitive are the spectral fits even to individual measurements. Those two values are definitely statistically different, but one has to note that the added measurement is from over 35 years ago while the remaining ones are presumably from the last few years. In such cases the inhomogeneity effects (see the Introduction) that are able to change the pulse broadening over time may have played a significant role. This case strongly shows how such effects may influence the measured spectral slopes.

Another pulsar published already in Paper~1 was PSR~B1641$-$45. For this one the previous spectral slope was 3.81 and was based on two frequency observations only (hence no error estimate). In our current analysis, where we also added observation from  \citet{rickett77}  the resulting $\alpha$ is 3.84$\pm$0.08, hence in this case the 35-year old measurement appears to be in a very good agreement with the newer data. Finally the last case of the pulsar common for Paper~1 and our current analysis is PSR~B1758$-$23, the source with the highest DM value (1073.9 pc cm$^{-3}$). As a result of our previous analysis we quoted the value of $\alpha=4.92\pm0.11$, while in our current approach we added new data obtained for profiles from the ATNF Database, as well as added an archival measurement from Manchester~et~al.~(1985). The addition of the new data points clearly showed that the lower-frequency $\tau_d$ measurements show a completely different slope than the high frequency measurements at 2.6 and 4.8 GHz. These high frequency data were included in the previous fit (which consisted of 4 points in total), but in our current analysis we had 10 measurements available (see the online appendix) so we were able to exclude the two suspicious high frequency ones. As a result the obtained $\alpha$ value is much lower and the pulse broadening slope is even even flatter than the results published earlier by L01 ($3.91\pm0.4$, based on three frequencies).

\subsection{Scatter time scaling index versus the dispersion measure}

\begin{figure}
\resizebox{\hsize}{!}{\includegraphics{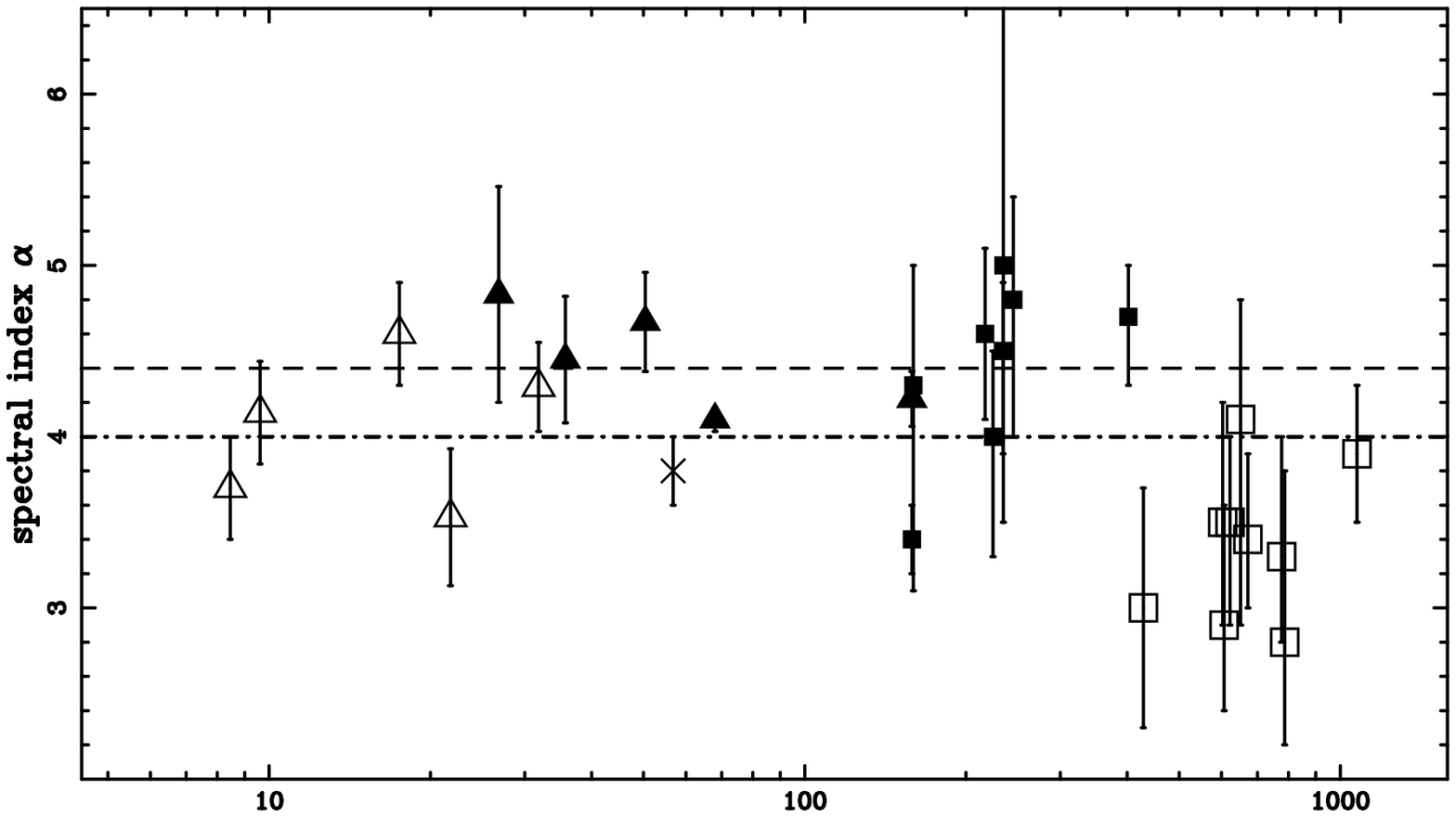}}
\vskip1mm
\resizebox{\hsize}{!}{\includegraphics{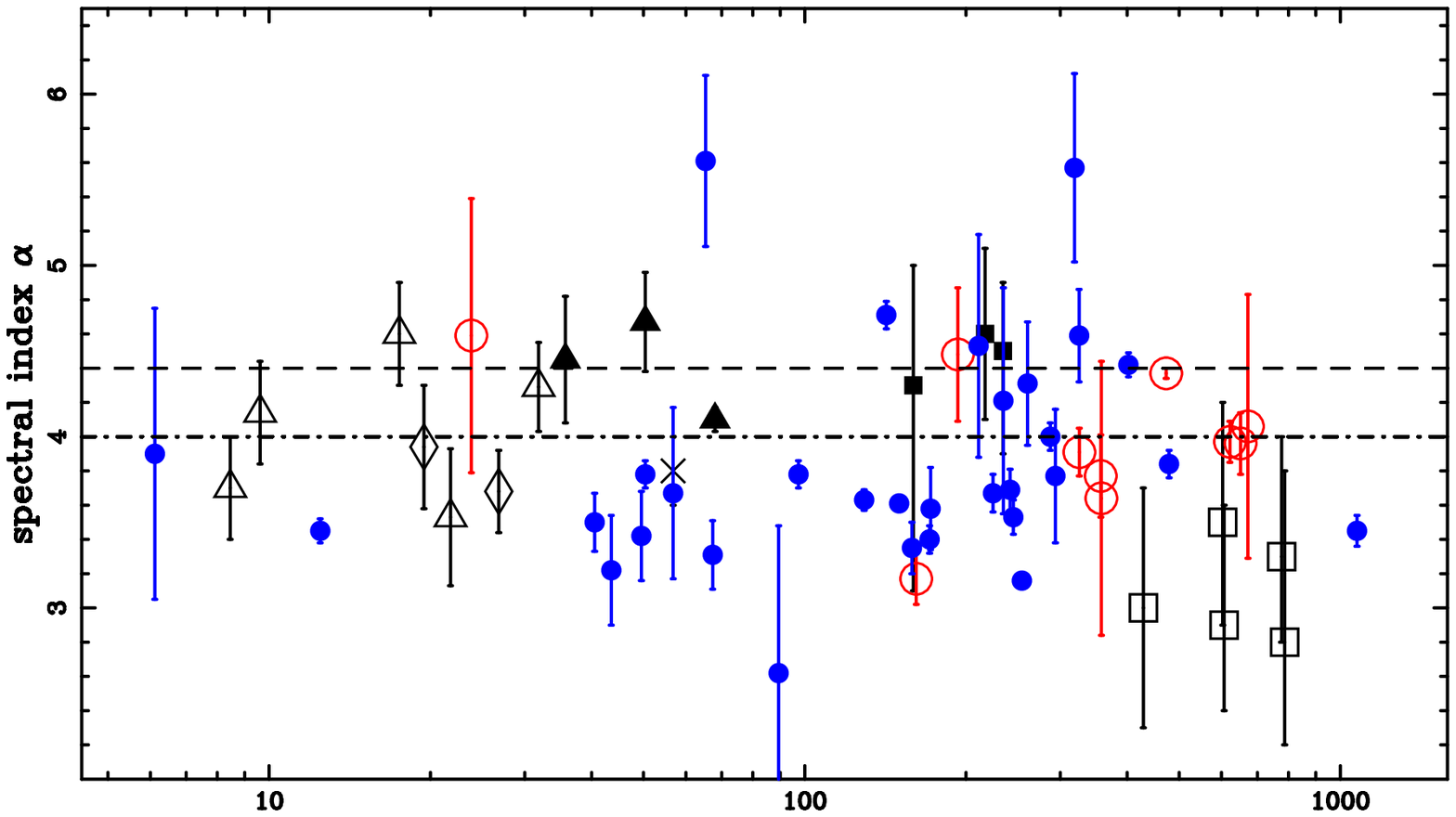}}
\vskip1mm
\resizebox{\hsize}{!}{\includegraphics{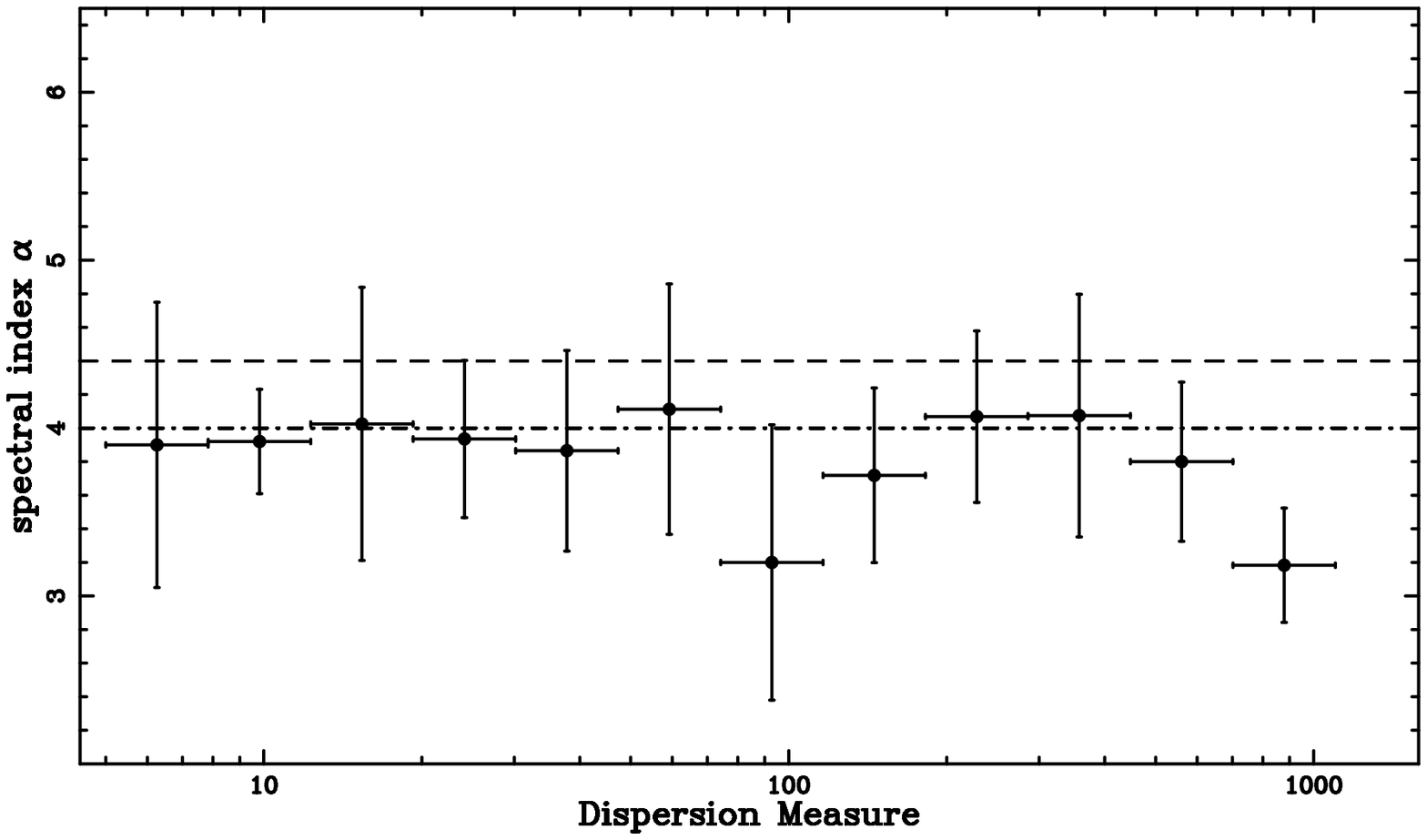}}

\caption{A plot of the spectral index of pulse broadening $\alpha$ versus the dispersion measure. Top panel shows the data from L04 only, the middle panel includes our measurements. Empty squares represent the measurements from L01, filled squares - L04, empty triangles - \citet{johnston98}, filled triangles - \citet{cordes85}, cross - \citet{kuzmin02}, diamond - \citet{lewan11, daszuta13}, open circles (red) - \citet{lewan13}, full circles (blue) - this paper. The bottom panel shows the averaged alpha values (for 12 bins in $\log$DM).\label{alpha_dm}}
\end{figure}

Figure~\ref{alpha_dm} shows the values of the scatter time spectral index $\alpha$ versus the dispersion measure. For comparison we are showing two plots: the re-created plot of L04 (top) and the current plot of the spectral slopes for 61 pulsars (L04 had 27 sources). The plot shows only the values of $\alpha$ for pulsars that had $\tau_d$ measured on at least three frequencies (in this regard it is different from the respective plot we showed in Paper~1), and were not excluded for any of the reasons stated in above paragraphs. To create the new version of the plot in addition to the data used by L04 and all the references therein (see the figure caption for a full reference list) we used our data presented in Paper~1 (these are shown as open red circles) and the results of our current research (full blue circles). We also added two $\alpha$ estimations from our scintillation studies of PSR~B0329+54 \citep[$\alpha=3.86\pm0.24$,][]{lewan11} and PSR~B0823+26 \citep[$\alpha=3.94\pm0.36$,][]{daszuta13}; these are shown in Figure~\ref{alpha_dm} by diamonds.

The current version of the $\alpha$~versus~DM plot looks significantly different from the one showed by L04, and not only because we increased the number of measurements (by the factor of 2.25). First off one has to note that some of the L01 and L04 measurements had disappeared, as we replaced them with our own, updated values for the given sources.  In most cases the new values agree with the theoretical predictions better, which in fact counters the L01 argument that most of the high-DM pulsars show significant deviation from theoretical predictions. That is true for L01 measurements (open squares) for high DM pulsars, as we already showed in Paper~1 at least in two cases we replaced measurements which were way below theoretical predictions (i.e. $4.0<\alpha<4.4$) - this can be seen as two of the open squares from around DM=600 on the top plot being replaced by two open circles with $\alpha$ very close to 4.0 on the new version of the plot. As one can see for 7 pulsars from L01 which had very low values of $\alpha$ there is only 5 remaining. While the 2 (out of 7) new  ``theory consistent'' values of $\alpha$ may not seem as much let us point out that these were the only two pulsars for which we had new data, for the rest we had to retain the $\alpha$ measurement from L01. 
 
Similarly, for the mid-DM pulsars analysed in L04 there was a few for which the values of $\alpha$ were greater than 4.4 (although consistent with thin screen predictions within error estimates).  Our current analysis for some of these pulsars 
yielded more theory-compliant $\alpha$ values, and were replaced by full circles on the middle plot.


Another interesting result is that the values of $\alpha$ we got from the scattering analysis when compared to the estimations from decorrelation bandwidth seem to show similar spread,  when we limit the range to DM$<300$~pc~cm$^{-3}$.  In the middle plot of Fig~\ref{alpha_dm} the scattering-based measurements are shown as circles and squares, whereas the scintillation-based estimates are represented by triangles, diamonds and a cross. In this DM range in cases of both types of measurements there is a solid number of pulsars for which $\alpha$ agrees with the theory-predicted range ($4.0<\alpha<4.4$), and the majority of deviations is towards lower values of $\alpha$, but usually not much lower than $\alpha=3.5$. 

The general idea of L01 and L04 when interpreting the $\alpha$ versus DM plot was, that while for low-DM pulsars the scatter time spectral index seems to agree with theoretical predictions, the pulsars with DM$>300$~pc~cm$^{-3}$  deviate from these predictions significantly. In our Paper~1 we agreed with this , however we postulated that the division is still there, however it appears at slightly lower DM values, around DM$=230$ to 250~pc~cm$^{-3}$. Looking at the current plot one may comment, that both the original, as well as our conclusion were based on data biased  by a small number of scatter time measurements.

The bottom plot in Fig.~\ref{alpha_dm} shows the averaged values of the scatter time frequency scaling index, and the averaging was performed for 12 $\log$-DM bins (in a DM range between 5 and 1100~pc~cm$^{-3}$). As one can see the averaged $\alpha$ values for the most part stay very close to the theoretically allowed value of $\alpha=4.0$. The only notable exceptions happen for  the bin around DM=100~pc~cm$^{-3}$ (which probably should be disregarded due to selection effects, as there are only two pulsars in that bin, and one of them has a suspiciously low value of $\alpha$), the last bin containing the highest DM pulsars (DM$>700$~pc~cm$^{-3}$) and possibly the lesser extent the bin before the last. There is no proof for the global deviation of $\alpha$ for pulsars with DM$>300$~pc~cm$^{-3}$ that L04 suggested. In the averaged $\alpha$ values significant deviations from the theoretical predictions seem to start at much higher values of DM. Even for the DM-bin covering the 500-600~pc~cm$^{-3}$ range the average $\alpha$ value is statistically (within the 1-$\sigma$ shown on the plot) in agreement with the theoretical predictions. As for the last DM-bin (DM$>700$) we have to point that the deviation may be possibly be due small number statistics, since its average $\alpha$ value is based on only three pulsars.

\begin{figure}
\resizebox{\hsize}{!}{\includegraphics{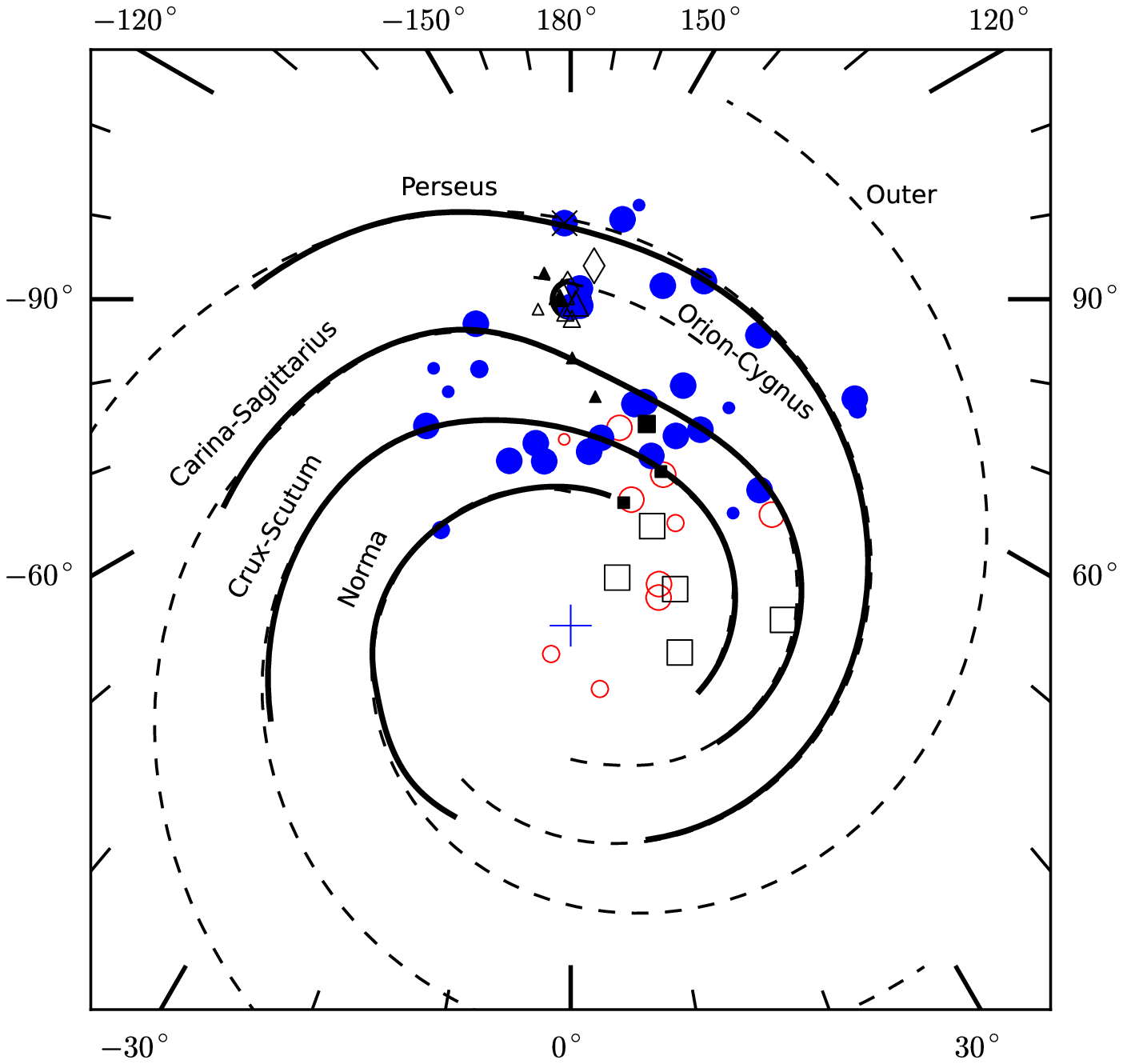}}
\vskip-3mm
\caption{Pulsars with measured scattering spectral index in the galactic X-Y coordinates with the spiral arms positions based on NE2001 model \citep{cordes2002}. Symbol (and colour) coding the same as in Fig.\ref{alpha_dm}. The size of the marker indicates the deviations from theoretical predictions. The largest markets represent pulsars with the $\alpha$ values  below 4.0, the smallest ones - objects with $\alpha>4.4$.  \label{alpha_xy}}
\end{figure}

While the averaged values seem to be largely consistent with the simple scattering theory, for the individual pulsars large deviations are much more common.  Apparently these start already at very low values of dispersion measure, and continue over the whole range of available DMs. If anything it is only the range of the deviation that seems to be increasing for the larger DMs. With 60 pulsars in the plot the selection effects will still be pretty strong, but our interpretation is that with the increasing dispersion measure (and thus the distance) the probability of deviation also increases, but even amongst the highest-DM pulsars there are some pulsars for which the agreement with theory is reasonably good.

We also plotted the pulsars from out sample in the Galactic X-Y coordinates in Fig.~\ref{alpha_xy}. On this plot we present the positions of the pulsars, and the size of the graph marker represents the value of the pulse broadening spectral index. The medium sized markers were used for sources for which their $4.0<\alpha<4.4$ (within error estimates), the largest markers represent objects with $\alpha$ definitely lower than 4.0, and the smallest ones - pulsars with $\alpha>4.4$ (which is only 3 sources, corresponding to 3 smallest size full circles). 

We can safely state that there is no correlation between the value of the pulse broadening spectral slope and the distance and/or the position of the source in our Galaxy. One thing to remember when interpreting this plot is that the pulsar positions, especially for the more distant objects were based mainly on the galactic electron density model (NE2001, \citealt{taylor93}), and in many cases the precision of such distance estimation, and hence the precision of the actual pulsar position in the Milky Way is relatively low.

\subsection{Search for the evidence of inner scale effects}

As in Paper~1 we also searched our data for any evidence of the inner scale effects influencing the $\tau_d$ measurements in our database. Table~\ref{tab_inner} shows the results. In case of a few selected pulsars, for which we had a large number of $\tau_d$ measurements we split the frequency range into two sections. The frequency at which the division happened was different for all the sources, and decided upon the view of the data (see Fig.~\ref{plot_tau_vs_freq}). In selected cases we used $\tau_d$ measurements that were previously excluded from the overall analysis.

\begin{table}
\caption{Pulse broadening spectral slopes calculated in different frequency ranges for selected pulsars from our database. $\alpha_{low}$ is the spectral index in the low-frequency range, $\alpha_{high}$ - in the high frequency range and $\nu_{div}$ denotes the frequency at which the data was divided.\label{tab_inner}}
\begin{tabular}{llcl}
\hline
Pulsar & $\alpha_{low}$ & $\nu_{div}$(MHz) & $\alpha_{high}$ \\ \hline
B0809+74 & 1.51$\pm$0.28 & 44& 3.90$\pm$0.85 \\
B1154$-$62 & 3.79 & 685 & 5.24$\pm$ 0.07 \\ 
B1557$-$50 & 4.28$\pm$0.44 & 900 & 3.76$\pm$0.41 \\
B1758$-$23 & 3.45$\pm$0.09 & 1600 & 4.76$\pm$0.62 \\
B1821$-$19 & 5.2$\pm$1.9 & 500& 3.86 \\
B1859+03 & 4.42$\pm$0.07 & 925 & 6.0$\pm$.0.6 \\
B1913+10 & 3.66$\pm$0.13 & 925 & 4.47 \\
B1946+35 & 3.62$\pm$0.06 & 430 & 4.00 \\
\hline
\end{tabular}
\end{table}

As it was mentioned earlier in the Introduction, the inner scale effects should manifest themselves as a flattening of the spectral slope at low observing frequencies. One thing to mention is, that the frequency at which it should happen actually depends on the inner scale value, and it is impossible to know it beforehand. 

Again, as it was in case of our similar analysis in Paper~1, the results presented in Table~\ref{tab_inner} are  highly inconclusive, mainly due to the quality of the data and subsequently - the reliability of the scaling index fits, which are often based upon just two measurements. For instance, in the cases of B0809+74, B1154$-$62 and B1859+03 one of the slope measurements is highly unrealistic. For B1557$-$50 (similarly to what we noticed in Paper~1)  and B1821$-$19 the lower frequency slopes are actually greater than the high frequency ones; an opposite to what we would expect from the inner scale effects. In the remaining cases - B1758$-$23, B1913+10 and B1946+35 - we got the desired effect, but especially the high frequency slopes are highly uncertain - in case of the first pulsar there is a large error bar (fit based on just 3 widely spread data points) and in case of the other two objects we have  error-less values based on just two frequencies. Also, one has to note, that for  a number of sources  shown in Table~\ref{tab_inner} the slope differences are not significant statistically (usually below 3-$\sigma$ level).

The above clearly shows that to perform this kind of analysis one needs much better quality data, especially when it comes to the frequency coverage of the pulse broadening measurements. Also, such data should be free of the inhomogeneity and anisotropy effects, preferably the measurements should be performed quasi-simultaneously at all the frequencies, at least in the sense of the inhomogeneity scale which is presumed to be of the order of a few weeks \citep{rickett09}.

\section{The strength of scattering}

In Table~\ref{table_alpha} one can find the estimated fluctuation strength for the pulsar's line-of-sight $C^2_{n_e}$.   Following L04 (see their Eq.6) we calculated this value using the parameters of the fit that yielded our scatter time spectral slope $\alpha$. Using these parameters we  extrapolated the scatter time to a uniform frequency of 1~GHz. Next, using the value of $C_1 = 1.16$ and Equation~\ref{def_c} we calculated the expected decorrelation bandwidth at this frequency, which we entered into L04's Eq.~6 to obtain the value of $C^2_{n_e}$. 

In most cases the results are much higher than the canonical value expected for homogeneous medium ($\log C^2_{n_e} = -3.5$, see \citealt{johnston98}), and this was also reported earlier by other authors (see for example L04).  The lowest value we found was for PSR~B0809+74 with $\log C^2_{n_e}$ = -4.72 which is not uncommon for nearby pulsars (and this source has the lowest DM in our sample). The highest value we found was $\log C^2_{n_e} = 2.33$ for B1758$-$23 - the highest DM pulsar. When looking at the values of the fluctuation strength one has to remember that the estimation is very sensitive to the value of the scatter time spectral index, especially if one underestimates $\alpha$ the values of  $C^2_{n_e}$ may be overestimated. Hence in cases where we are not certain of $\alpha$ (see previous paragraphs) in Table~\ref{table_alpha} we quoted the $\log C^2_{n_e}$ in brackets.

\begin{figure}
\resizebox{\hsize}{!}{\includegraphics{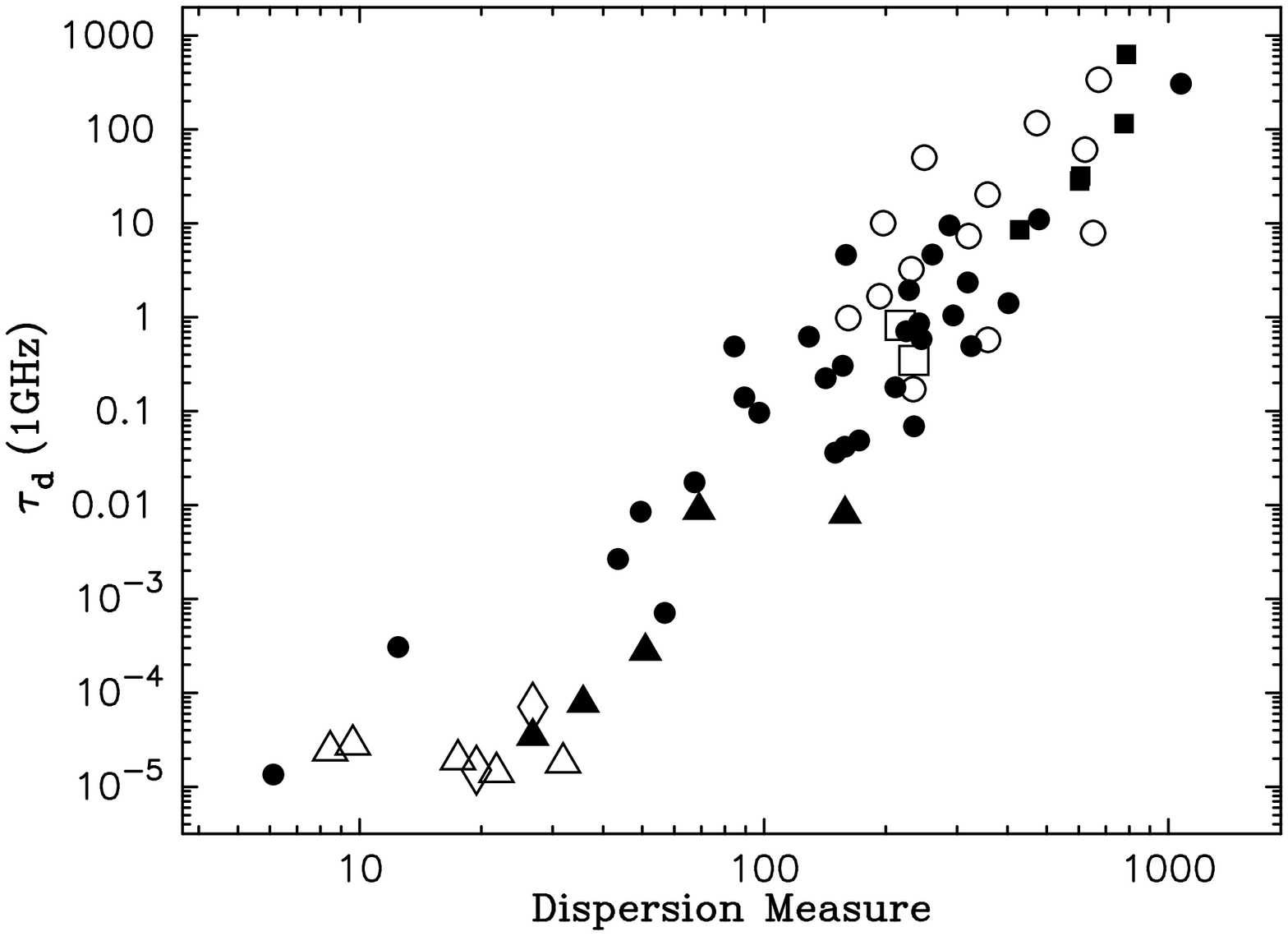}}
\vskip1mm
\resizebox{\hsize}{!}{\includegraphics{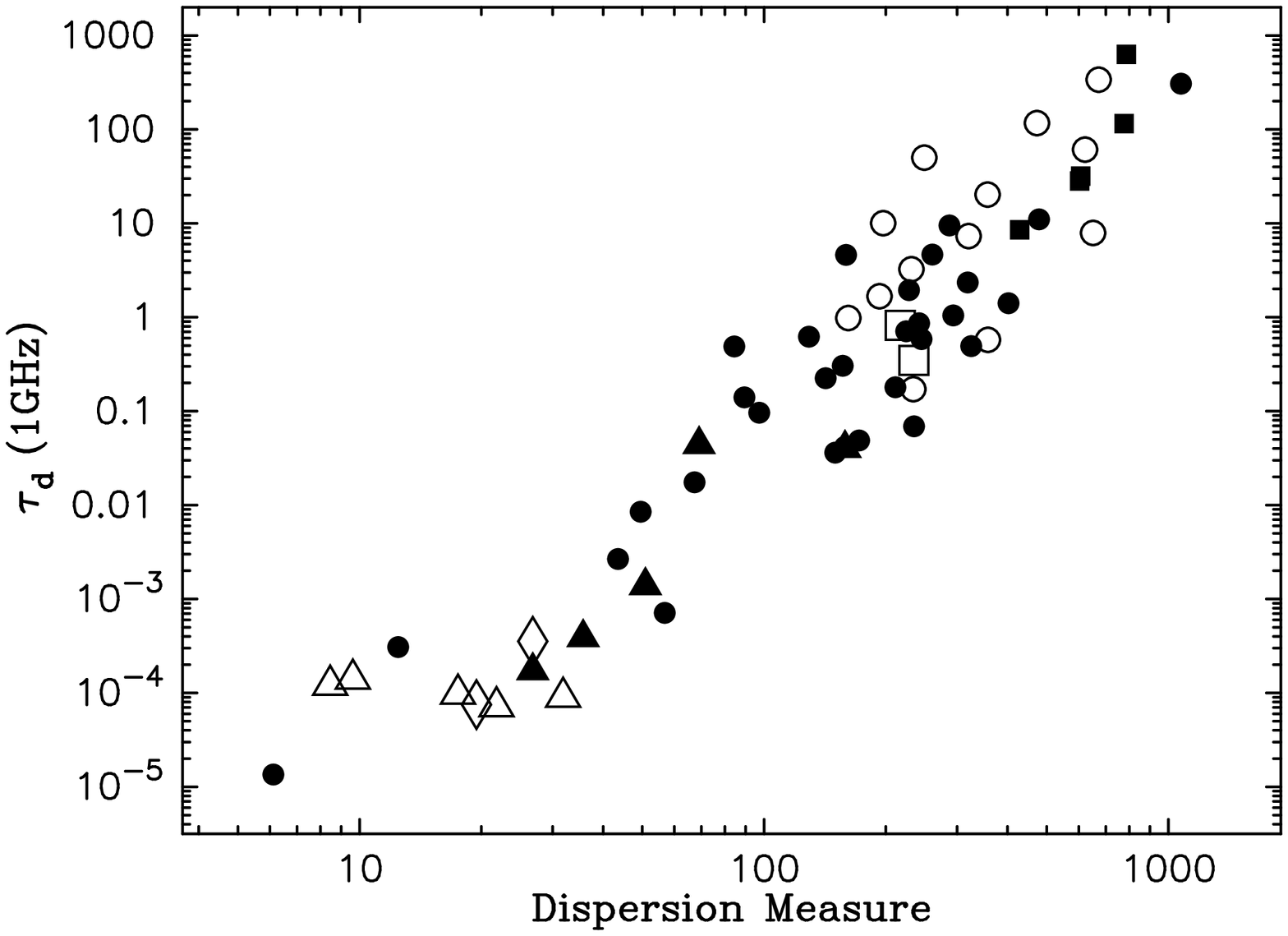}}
\vskip1mm
\resizebox{\hsize}{!}{\includegraphics{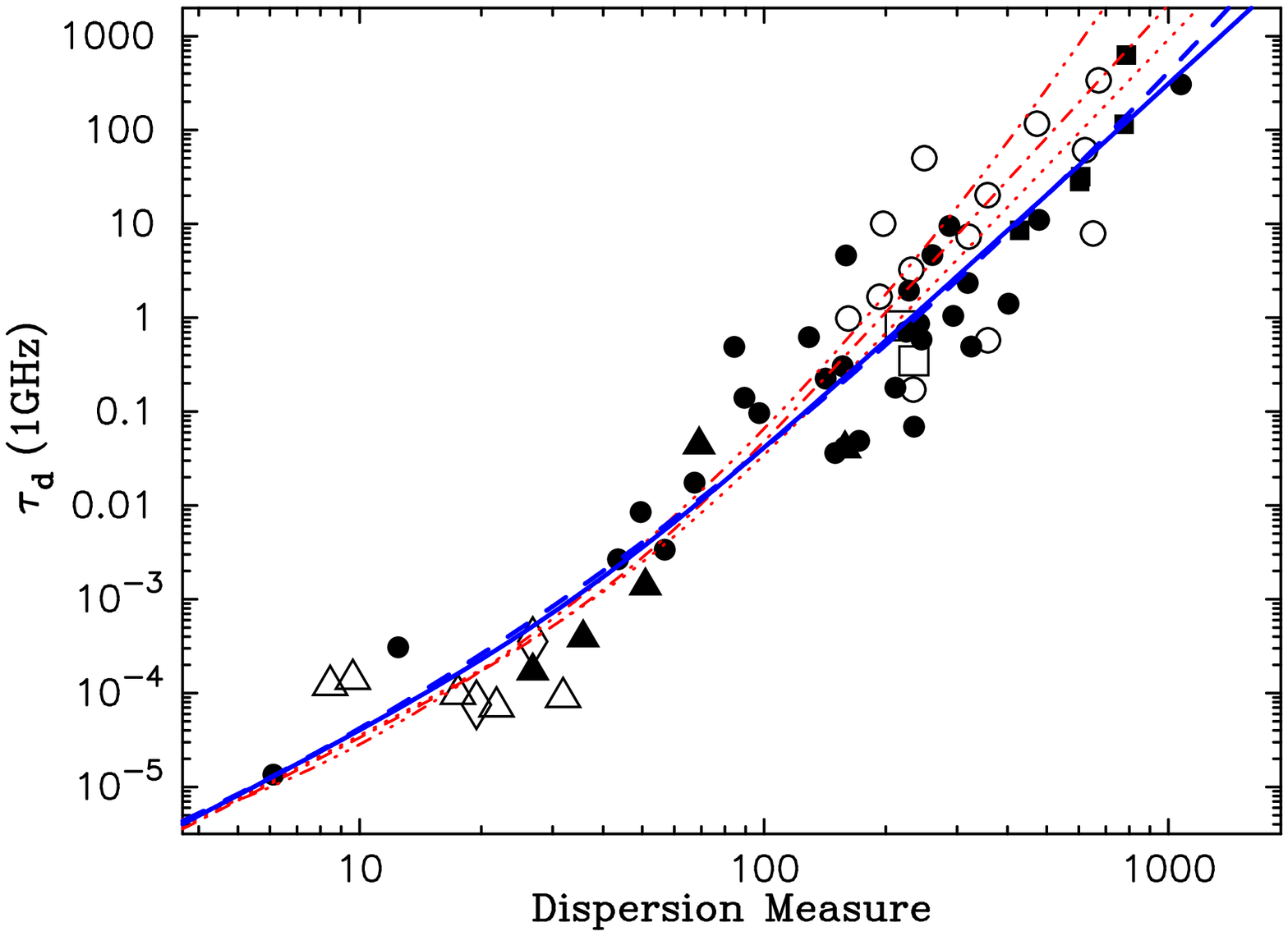}}
\caption{Three plots of the 1~GHz scattering versus the dispersion measure for two values of the $C_1$ constant: $C_1=1$ (top panel) and $C_1=5$ (middle and bottom panels). The scattering-based estimates are shown by empty squares - L01, filled squares - L04, open circles - \citet{lewan13}, full circles - this paper. The scintillation-based estimates are shown as: empty triangles - \citet{johnston98}, filled triangles - \citet{cordes85}, cross - \citet{kuzmin02}, and diamonds - \citet{lewan11, daszuta13}. The bottom panel also includes the lines corresponding to our fits  (the blue solid and dashed lines) as well as the previously proposed $\tau_d$ versus DM relations: the red dash-dotted line for \citet{rama97}, the dotted red line for L04 and the three-dot-dashed line for \citet{bhat04}. \label{bhat}}
\end{figure}

The value of the fluctuation strength $C^2_{n_e}$ is one of the ways to describe the strength of scattering happening along the line-of-sight of an object, but by definition it is supposed to be distance independent. To estimate the actual amount to scattering we decided to follow the idea of \citet{bhat04} and also L04. To compare the measured values of pulse broadening between different sources we re-scaled the scattering of the pulsars from our sample to a standard frequency of 1~GHz, using the actual frequency scaling  power-law, i.e. the values of $\alpha$ and $b$ we got from our fits (see Eq.~\ref{fit_form}). We also re-calculated the values of $\tau_d$ for the frequency of 1~GHz for the data we published in Paper~1, as well as for all the data from the references used  to create Fig.~\ref{alpha_dm} (see the figure caption for the reference list), making our own fits to the scatter time or decorrelation bandwidth data where necessary. For the scintillation data we interpolated (or extrapolated) the value of the decorrelation bandwidth to the observing frequency of 1~GHz, and then calculated the corresponding pulse broadening time at this frequency using formula from Eq.~\ref{def_c}, assuming (initially) the value of $C_1=1$. The result is shown in the top panel of Figure~\ref{bhat} . For the purposes of this plot we used only objects in which we are certain about the values of the scatter time scaling index $\alpha$ (i.e. the values of $\tau_d$ at 1~GHz that in Tab.~\ref{table_alpha} are not quoted in brackets).

One thing to notice about this plot is the apparent discrepancy between the scattering-based values of $\tau_d$ at 1~GHz and the scintillation based estimates. The latter are presented as diamonds and both types of triangles, while the remaining markers (full and empty circles and squares) are from scattering measurements. The discrepancy is visible especially for the $\tau_d$ estimations based upon data from \citet{cordes85} (full triangles). It seems that the bunch of scattering measurements does not follow the trend defined by these data, and the scattering estimates are slightly higher on the plot or the scintillation estimates are too low. Such assumption may be of course biased by a relatively poor overlap between these two datasets, and the apparent discrepancy is based only on a very few estimates, however we believe it to be true.

To asses the discrepancy we looked at the pulsars for which we were able to get scatter time at 1~GHz estimates from both the scattering and the scintillation observations, and we found 4 such sources. We had to exclude 2 of them from start, since we disregarded our $\alpha$ values for B0329+54 and B0823+26. In the remaining two we have the Vela pulsar (B0833$-$45) for which our estimate from scattering measurements was  0.0466~ms, while based on decorrelation bandwidth measurement from \citet{cordes85} we got $\tau_d$(1 GHz)$ = 0.00804$~ms which means difference by a ratio of 5.8. The other one is B1933+16 for which we obtained the standard frequency $\tau_d$ of 0.0417 from scattering and 0.00883 from scintillation data (again based on \citealt{cordes85}) - a ratio\footnote{for the less reliable ratio estimates we got a ratio of 9.7 for B0329+54 and a ratio of 56 in case of B0823+26, but for both pulsars the values of $\tau_d$ at 1~GHz from scattering are hugely overestimated due to a very probable underestimation of $\alpha$} of 4.7.

The difference between these values can be explained assuming that the $C_1$ constant is in reality not close to the unity, but rather close to 5. We explored this hypothesis and used the value of $C_1 = 5$ to create the plot shown in the middle panel of Figure~\ref{bhat}.  One has to remember however that this value is based on just two pulsars, and the scatter time and scintillation measurements were not performed simultaneously. Also, our way of estimating $C_1$ by comparing the decorrelation bandwidth estimate and the scatter time estimate after extrapolating them both to the frequency of 1~GHz may be a source of additional error due to the uncertainties in the scaling indices. Since the scaling indices for both of these quantities are not the same - the resulting $C_1$ value will depend on the choice of the standard frequency. It would be probably to far to suggest at this point that the value of $C_1$ may be frequency dependant (that would require us to re-evaluate the whole scattering and scintillation theory), but in our approach that would be an expected outcome.

Finally, the pulsars we used are in some regards fairly non-standard: the Vela pulsar profile shape makes it extremely hard to obtain a reliable scatter time measurement (the intrinsic pulsar profile apparently exhibits a scatter-like tail), whereas PSR B1933+16 shows an extremely low scaling index ($\alpha=3.35$) and the peculiarity of its line-of-sight was discussed by many authors before (see L04 for example).

The value of $C_1$ we suggest here is also much larger than unity which is in contradiction to what \citet{rickett77} noted, that there is very little flexibility about the $C_1$ value in the theory. Also, the value of $C_1=5$ is in contradiction with \citet{backer74} findings, as he obtained $C_1=1.07$, incidentally - for the Vela pulsar that we also used. Another evidence for $C_1$ close to unity was provided by \citet{cordes1990} in their analysis of the scintillation and scattering properties of the millisecond pulsar PSR~B1937+21. Taking into account all of the above we suggest that our result should be viewed with a dose of scepticism. Nevertheless, similarly to L04  we believe that the evidence for larger values of $C_1$ is getting stronger, as is the evidence that the value of $C_1$ may be line-of-sight dependant, and that the ISM is in fact highly anisotropic and/or inhomogeneous at least along some of the lines-of-sight.

We based our further analysis on assumption that indeed $C_1=5$, meaning that the middle plot in Fig.~\ref{bhat} is valid. The next step was to try to model a relation between $\tau_d$ at 1~GHz and the dispersion measure. First we used a relation proposed by \citet[][also used by L04]{rama97}: $\tau_d = A \ {\rm DM}^{\gamma}(1 + B \ {\rm DM}^{\zeta})$. Following L04 we fixed $\gamma=2.2$ (to estimate the deviations from Kolmogorov value), which yielded the best fit with:

\begin{equation}
\label{fit_l04}
\tau_d(\rm ms) = 2.26\times10^{-7} \ {\rm DM}^{2.2}(1+0.00205 \  {\rm DM}^{1.74}).
\end{equation}

The results of this fit are shown in the bottom panel of Fig.~\ref{bhat} as a solid blue line. The values of all the parameters are quite different from what L04 obtained, although one has to remember that we used data for 60 pulsars, while they had only 27 pulsars available, and also they assumed $C_1=1$. Especially the difference in the value of $\zeta$ is striking, as L04 got $\zeta=2.3$ and earlier \citet{rama97} estimated $\zeta=2.5$. This means that our data indicates a much flatter scatter time versus DM dependence, especially in the high-DM range.

Another form of this dependence we tried was the parabolic relation proposed by \citet{bhat04}; adopting this we got the best fit corresponding to a relation for scattering at 1~GHz:

\begin{equation}
\label{bhat2}
\log \tau_d({\rm ms}) = -6.344 + 1.467 \log{\rm DM} - 0.509 \  (\log{\rm DM})^2,
\end{equation}

\noindent
which is represented by a dashed blue line in the bottom panel of Fig.~\ref{bhat}.
As one can see we got much stronger linear term as for \citet{bhat04} it was equal to 0.154 (our value of 1.467 is almost 10 times greater), and much lower quadratic term; our 0.509 versus 1.07 from \citet{bhat04}. One has to remember however that in that paper authors used a different approach to the problem, as they used 371 individual scatter time estimates and for the purpose of frequency scaling they used the scaling index $\alpha$ as one of the fit parameters, and in the fit they assumed it to be the same for all the pulsars. Using this method they obtained the average value of $\alpha=3.86\pm0.16$, which can be considered  remarkably close to the average values from our data (see the bottom plot in Fig.~\ref{alpha_dm}). In our study we used the actual estimates of the frequency scaling indices for 60 pulsars (each pulsar is represented by a single mark on the plot) to obtain the standardized scatter time value for a given object. These indices were different for every pulsar (ranging from $\alpha=2.73$ to $5.61$) and they come from about 300 individual scatter time estimates, as each of our $\alpha$ (and thus the 1~GHz $\tau_d$) values  was based on at least 3, and up to 9 individual scatter time estimates. 

In the bottom plot if Fig~\ref{bhat} we are also showing the results of the previous attempts to find the $\tau_d$ versus DM relation. The red dash-dotted line shows the formula given by \citet{rama97}, the dotted line is based on the formula given by L04, and the three-dot-dashed red line is based upon the formula from \citet{bhat04}; all these lines were adjusted for the value of $C_1=5$. With this adjustment (which is supposed to work better for the low-DM range, where the scintillation based estimates dominate) one can see that all the proposed relations basically agree in the low- to mid-DM range, especially given the large spread of data. The discrepancies appearing at high-DM values are also probably  due to the $C_1$ adjustment, but one thing to note is that the formula given by L04 (dotted line) seems to be the most similar to our results and is showing a similar slope in the high-DM range. This should not come as a surprise since our dataset is just an extension of their work. 

The differences between our approach and the one that \citet{bhat04} used, which in turn influenced  the available pulsar sample, are probably responsible for the different results of the $\tau_d$ versus DM relation. In both cases however one has to 
note a large scatter present in the data.  The source of the scatter may be astrophysical in nature (like the strongly line-of-sight dependant nature of the scattering), as well as method-dependant, and probably both. In case of \citet{bhat04} one of the sources of an additional spread in the data may be the assumption of an uniform value of $\alpha$, what does not agree with observations. In our method the additional scatter may be present due to using different scaling indices, as in such case the method will yield different outcome depending on the value of the observing frequency to which we ``standardize'' the amount of scattering. Finally, both methods will suffer due to the uncertainties of individual pulse broadening measurements.

Because of the data scatter we were unable to perform any meaningful fit quality testing to check which of the proposed functional forms fits the data better. The values of normalised $\chi^2$ are similar for both models (of the order of 0.44) which does not allow to decide which of the fits is more reliable. Large scatter of the data also translates to very low precision of the estimated parameters. Also, it is quite possible that the relation has another form, which we did not consider.

Regardless one has to remember that any kind of relation derived from this type of analysis will be only an attempt to average scattering happening along sometimes vastly different lines-of-sight, therefore a large scatter will be always present. The relations between the pulse broadening and DM we propose here should be at best considered only a very rough estimates. For a given pulsar, when calculating the expected scatter time from any of these formulas, or the ones proposed earlier by L04, \citealt{rama97} or \citet{bhat04} one has to remember that the actual scattering may be at least one order of magnitude larger or smaller than the derived value, depending on the properties of ISM along the given line-of-sight.

Also one has to be aware that this kind of analysis will be always prone to errors due to various effects mentioned earlier in this paper, such as anisotropies, the inhomogeneity of ISM and our lack of knowledge about the distribution of free electrons along the line-of-sight.

\section{Discussion and Conclusions}

In this paper we performed analysis of the multi-frequency scatter time measurements, following L01 and L04 and focusing on pulsars for which the pulse broadening was estimated at multiple frequencies. To do so we constructed a database of all the scatter time and decorrelation bandwidth measurements we could find. This database was assembled from both our own observations, as well as the data available - both in form of pulse profiles as well as the data from the literature. It consists of over a 100 pulsars which had pulse broadening and/or the decorrelation bandwidth measured at multiple frequencies. Over the course of our analysis, and using similar datasets published previously we were able to gather a number of sources for which we were able to reliably ascertain the value of the pulse broadening frequency scaling indices which are based on scatter time measurements from at least 3 frequencies. The largest previous analysis of a similar dataset was done by L04 (see also references therein) who had the data for 27 objects. Based on our data published in Paper~1 we added the (three-frequency based) estimates of $\alpha$ for 11 pulsars and some of these were updates of the values published in L01 and L04. As a result of our current study we added or updated the $\alpha$ and $\tau_d$ (at the standard frequency of 1~GHz) values for 33 sources, which yielded a total of 60 pulsars with reliable estimates of the scattering frequency scaling indices, over two times the number L04 had.

A large portion of resulting values of the $\alpha$ indices does not conform with the predictions based on the most commonly used models of the scattering phenomenon. The simplest model of a scattering with a thin screen geometry  and Kolmogorov-type density fluctuation within the screen predicts the pulse broadening spectral slope of 4.4. Using other fluctuation models one can get expected values of $\alpha$ down to 4.0, and applying different geometries of ISM along the line-of-sight one can predict $\alpha$ values lower than 4.0.

 We extensively discussed the phenomena that influence the measured value of $\alpha$ in our previous paper (Paper~1, \citealt{lewan13}), and virtually all the arguments pointed there still hold. There are a few reasons for which the measured pulse broadening spectral slope may be lower.  The main is probably  geometry of ISM, including the possibility of scattering on multiple screens (see also L01 and L04). If the  actual spectrum of the fluctuation density different from Kolmogorov's (see \citealt{gupta00} for a brief summary) it will also play its role in the general picture.  Another source of discrepancy may be the inner scale effects \citep[see also][]{rickett09} which can affect and flatten the spectral slope below certain observing frequency. Finally, the anisotropy and the inhomogeneity of the turbulence may also influence the observed scatter time spectra, both directly, as well as indirectly - both of these effects may cause the scattering parameters to vary in time, hence they may make it impossible to create a proper spectra of pulse broadening.
 
 Our current analysis mainly due to its completeness (with regard to the previous similar attempts) allows us also to  study the dependence of basic scattering parameters with dispersion measure. As we mentioned above, prior to our work the most complete analysis of this kind was made by \citet{L01,L04}. 
 Based on the data they had available they postulated, that the scatter time spectral index $\alpha$ stays with good agreement with the theoretical predictions for pulsars with DM$<300$~pc~cm$^{-3}$, while for objects with larger DM the slope is much lower ($\alpha=3.48$ on average).
 
In Paper~1 we modified this conclusion only slightly, but with the new or updated estimates of $\alpha$ we obtained (especially for pulsars with low- and medium-DMs) it seems that there is no clear division between the low-DM and high-DM pulsars. The deviations from the theoretically allowed range of scatter time scaling indices ($4.0<\alpha<4.4$) start from the lowest DM values for selected sources. If anything, the range of the deviations seem to increase with higher dispersion measures. Also, one has to note that not all mid- and high-DM objects deviate from theoretical predictions, and it seems that it is only the probability of deterring from theory that increases for larger DMs.

Collecting $\alpha$ estimates for 60 pulsars allowed us to average the values over various DM intervals (see the bottom plot of Fig.~\ref{alpha_dm}). The result was that in a broad range of dispersion measures the average value of the spectral scaling indices stays almost the same and only slightly below $\alpha=4.0$, and that is despite the fact that for individual sources the value of the scaling index can vary a lot. Only for DM$>$500~pc~cm$^{-3}$ we noted a slight deviation of the average $\alpha$ value, but becomes statistically significant  for very large DMs ($>$700~pc~cm$^{-3}$). The average value of the scaling index  from all of our data is $\alpha=3.89$,  for pulsars with DM$<$500~pc~cm$^{-3}$ the average is $\alpha=3.95$ and for sources with DM$>$500~pc~cm$^{-3}$ the average value is $\alpha=3.49$ (based on 8 pulsars).

As it was previously known the simplistic thin-screen model can not explain the scattering properties for very distant objects. Especially with the possibility of multiple scatterings along the line-of-sight one could expect, that the number of such occurrences with increase for more distant pulsars,  however there is no guarantee for that, as most of the astrophysical objects that could play the role of a thin scattering screen (such as dense HII regions) can be relatively small. So, while the chance that the line of sight will cut such screen will increase with the distance, there may still be some selected directions in which that will not happen. The probability of undisturbed line-of-sight however should rapidly decrease with distance, and seemingly this is what we observe.
 
We also analysed the actual amount of scattering visible in pulsars from our sample versus the dispersion measure. Similarly to L04, and contrary to \citet{bhat04} for each individual source we used the actual scatter time frequency scaling law to estimate the amount of pulse broadening at the frequency of 1~GHz. The results are shown in Fig.~\ref{bhat}. To bind the scatter time estimates from the actual pulse broadening with the values obtained from decorrelation bandwidth measurements we decided to adjust the value of the $C_1$ constant which appears in the relation connecting these two quantities (see Eq.~\ref{def_c}). Based on measurements of of the Vela pulsar, and B1933+16 we used the value of $C_1=5$,  although one may argue that the evidence for such value is not very strong. To describe the $\tau_d$(1~GHz) versus DM dependence we tried to model it using the formulae proposed earlier by \citet{rama97} as well as the parabolic relation used by \citet{bhat04}. In both cases the resulting formulas are seemingly quite different from the ones published previously (when one considers the fitted parameters), however upon inspecting the plots they seem in agreement with each other, at least in the low- to mid-DM range. If anything, our  $\tau_d$ versus DM  relations look more linear (on the log-log plot) than in the previous attempts to recreate it. This also means that for the mid-DM pulsars we (on average) expect more scattering, while for high-DM objects the rise of the scattering strength will not be as sharp as it was previously predicted. 

Our database of multi-frequency scatter time measurements and the frequency scaling index estimates presents a significant extension of the previous similar attempts, there is however still a lot of possibility for improvement. The frequency coverage can be, and probably will be better in the nearest future, given the fact that the advances in technology will allow us to use wide-band receiver systems even at radio frequencies below 1~GHz (one of such examples is the GMRT upgrade which is currently happening). It will definitely be improved even more for the low-DM pulsars with the help of the low-frequency telescopes such as LOFAR. The future looks bright, which is really good, because - as our analysis presented here shows - we still do not know everything about the scattering of pulsar radiation in our Galaxy.

Finally, want to emphasize probably the most important outcome of our analysis: it appears that the properties of scattering of radio waves in the Milky Way are much more line-of-sight dependant than it was previously thought, even for more distant objects. Thus any attempts to generalize the behaviour of scattering may prove to be futile, and large scatter observed in  the scattering frequency scaling index, or the scattering strength, versus the dispersion measure plots will be always present.

\section*{Acknowledgments}
This paper was partially supported by the grant DEC-2012/05/B/ST9/03924 of the Polish National Science Centre. WL and JK also acknowledge the support of the grant DEC-2013/09/B/ST9/02177 of the Polish National Science Centre. MK is a scholar within Sub-measure 8.2.2 Regional Innovation Strategies,
Measure 8.2 Transfer of knowledge, Priority VIII Regional human resources
for the economy Human Capital Operational Programme co-financed by European
Social Fund and state budget. We are grateful to the anonymous reviewers for their help on improving our paper. Additional thanks for Marta Dembska and Krzysztof Maciesiak for technical help.

\appendix

\setcounter{table}{2}
\clearpage
\section*{Appendix: Multi frequency scatter time measurements database}

In this appendix we present the results of the scatter time measurements, both our own as well as the ones found in the literature, that were included our multi-frequency scatter time database, along with the data published earlier in Lewandowski et al. (2013). We present the data for all the pulsars from our studies for which we obtained new measurements, either from our own observations or from the archived profiles.

\vskip5mm
Below is the list of references to the table:
\begin{description}
\item (1) Alukar, Bobra \& Slee (1986)
\item(2) our analysis based on the profiles from the EPN Database
\item(3) Rickett (1977)
\item(4) Manchester, D'Amico \& Tuohy (1985)
\item(5)  Rankin (1981)
\item(6) Cordes, Weisberg \& Boriakoff (1985)
\item(7) Gullahorn \& Rankin (1978)
\item(8) Slee, Dulk \& Otrupcek (1980)
\item(9) Lewandowski et al. (2013)
\item(10) L{\"o}hmer~et~al. (2001)
\item(11) Ables, Komesaroff \& Hamilton (1970)
\item(12) our analysis based on the pulsar profiles from the  ATNF Parkes Database
\item(13) Kuzmin \& Losovsky (2007)
\item(14) our analysis of new observations performed at GMRT
\item(15) Staelin \& Shutton (1970)
\item(16) Ulyanov, Seredkina \& Shevtsova (2012)
\end{description}

\setcounter{table}{2}

\begin{table}
\caption{Full listing of the scatter time measurements\label{tau_sc_table}}
\begin{tabular}{lrlc}\hline
Pulsar & $f_{\rm obs}$[MHz] & $\tau_d$ [ms] & Reference$^*$ \\
\hline
\hline
B0136+57	&102&	25	6&	13\\	
	&408	&1.43$\pm$0.1	&2		\\
	&610	&1.23$\pm$0.04&	2	\\
	&925	&1.5$\pm$0.08	&2		\\
	&1408	&1.55$\pm$0.05	&2		\\
	&1642	&1.56$\pm$0.88&	2\\
	&&&\\
B0226+70&	111&	13$\pm$2&	13\\
	&1408	&0.3	&	2		\\
	&1420	&1.21$\pm$0.57&	2\\
	&&&\\
B0320+39&	40&	130$\pm$40&	13\\	
	&102&	0.4	&	2\\		
	&408&	1.02	&	2	\\	
	&	&1.44$\pm$0.14	&2		\\
	&	&0.39		&2		\\
	&	&0.5		&2		\\
	&610&	0.47	&	2\\		
	&	&0.95		&2		\\
	&	&0.43$\pm$0.04	&2		\\
	&	&4.39		&2		\\
	&	&4.92		&2		\\
	&910&	15.72$\pm$3.51&	2\\		
	&1400&	2.68$\pm$0.42	&2	\\	
	&1408&	20.48$\pm$1.52	&2		\\
	&4850&	22.28$\pm$7.55	&2	\\
	&&&\\
B0339+53&111	&25$\pm$5&	13\\		
	&606	&1.45$\pm$0.64&	2	\\	
	&	&1.54	&2	\\	
	&1400	&0.0057$\pm$0.001&	2\\
	&&&\\
B0353+52&	111&	45$\pm$10	&13\\
	&1420	&3.32$\pm$0.05&	2\\
	\\
	B0402+61&	111&	28$\pm$6	&13\\
	&400	&12.36$\pm$1.01&	2		\\
	&800	&0.19$\pm$0.11	&2		\\
	&925	&0.16$\pm$0.07	&2		\\
	&1330&	0.31$\pm$0.0002&	2		\\
	&1408&	0.11&2\\
\\
B0450$-$18 & 111 & 5.5$\pm$3 & 13 \\
                     & 150 & 5.8$\pm$1,6	&	1 \\
                     & 408	& 0.57 & 		2 \\
                     & 610	& 0.52$\pm$0.22	& 	2 \\
                     & 925	& 0.473$\pm$0.01		 &2\\
                     & 1404	& 0.27$\pm$0.06	&	2\\
                     &1642 &	0.29 &		2 \\
\multicolumn{4}{c}{{\it (continued on following pages)}}\\
	                       \hline 
\multicolumn{4}{c}{$^*$ full list of references at the beginning of the Appendix}
\end{tabular}
\end{table}

\begin{table}
\centering{\bf Table~\ref{tau_sc_table}.} Scatter time measurements (continued).
\vskip1mm
\begin{tabular}{lrlc}\hline
Pulsar & $f_{\rm obs}$[MHz] & $\tau_d$ [ms] & Reference$^*$ \\	\hline \hline
B0531+21&	44&	300$\pm$80&	13\\	
	&63&	100$\pm$20	&13		\\
	&111&	10$\pm$3	&13		\\
	&318&	0.23&		15\\			
	&2695&	0.002&		2\\
	\\
J0533+0402 & 102 &70$\pm$30 & 13\\
	        &370& 6.6$\pm$1.5& 2\\
	        \\
B0540+32&	111	&15$\pm$5&	13	\\
	&610	&3.8$\pm$0.08&	2\\
	\\
B0559$-$05&	111&	28$\pm$10&	13\\
	&400&	6.68$\pm$0.8&	2\\
	&800&	6.98$\pm$0.56&	2	\\
	\\
B0611+22&	111&	40$\pm$10	&13	\\
	&925&	2.37$\pm$0.23&	2	\\	
	&1408&	0.98$\pm$0.2	&2	\\	
	&1420&	0.21$\pm$0.09	&2	\\	
	&4850&	0.23	&	2\\
	B0621$-$04&	111&	45$\pm$4&	13\\
	&610&	1.95$\pm$0.92&	2	  	\\
	&&	1.99	0.93	&2	\\
	&1400&	1.74$\pm$0.27&	2\\		
	&&	1.76$\pm$0.27	&2		\\
	&1408&	1.71$\pm$0.59&	2\\
	\\
B0626+24 &111&55$\pm$30	&13\\		
 	&1408&	0.99$\pm$0.2&	2\\
 	\\
B0740$-$28     & 102	& 22$\pm$5.3		&13 \\
                     & 160	& 24.5$\pm$2.8	&	1\\
                     & 	& 21$\pm$3	&	8 \\
                     & 408	&5.34$\pm$0.39	&	2\\
                     & 422 &	1.09$\pm$0.2 &	2\\
                     &660	& 1.41$\pm$0.1	&	12 \\
                     &1390	& 1.16$\pm$0.17	&	12\\
                     &1518 &0.05$\pm$0.01 &	12\\
                     \\
B0808$-$47	&410& 33$\pm$1.5&		1	\\
	 & 660	&7.18$\pm$0.33&		2	\\	
	&1326.75&	1.94$\pm$0.7&		12 \\		
	&1382.5	&1.46	&	2	\\
	&1440	&5.14$\pm$0.94	&	2 \\
                      \hline 
\multicolumn{4}{c}{$^*$ full list of references at the beginning of the Appendix}
\end{tabular}
\end{table}

\begin{table}
\centering{\bf Table~\ref{tau_sc_table}.} Scatter time measurements (continued).
\vskip1mm\begin{tabular}{lrlc}\hline
Pulsar & $f_{\rm obs}$[MHz] & $\tau_d$ [ms] & Reference$^*$  \\
\hline
\hline
B0809+74	&23&	3&		16 \\	
	& &	10	&	16		\\
	&40&	4	&	16		\\
	&44&	3.9$\pm$1	&13		\\
	&62&	0.5		&16		\\
	&63&	0.6	0.2&	13		\\
	&102&	0.2	&	16		\\
	&410&	12.83$\pm$0.44&	2		\\
	&606&	8.37$\pm$0.67	&2		\\
	&925&	0.05$\pm$0.01	&2		\\
	&1408&	0.63$\pm$0.02	&2\\
	\\
B0823+26	&44&	12$\pm$3&		13\\	
	&63	&4$\pm$2	&	13\\		
	&400	&20	&		3	\\	
	&408	&0.23$\pm$0.02&		2	\\	
	&1055	&0.009$\pm$0.001&		2	\\	
	\\
B0833$-$45&	160	&621$\pm$109&		1\\	
	& 	&170$\pm$30	&	8\\
	&250	&20 &			11\\	
	&300	&9.4	&		3\\	
	&        & 12   &       11\\
	&318	&6				&2 \\
	&408	&4.7	&		11 \\		
	&636	&2		&	11	 	\\
	&1392.5&	2.25$\pm$0.08	&	2	 \\	
	\\
	&1420&	1.12&			11\\	
B0835$-$41&	160&	24$\pm$5.5&		1	\\
	&	&30$\pm$6	&	8		\\
	&685	&0.86$\pm$0.08&		12\\		
	&1590	&0.83$\pm$0.87&		12\\
	\\
B0839$-$53	&410&	20$\pm$2.4	&	1\\	
	&660&	2.14$\pm$0.24&		2\\		
	&1440&	0.24$\pm$3.8&		2\\
	\\
B0919+06	&44&	25$\pm$15&		13\\	
	&63&	9$\pm$5	&	13	\\
	&80&	15$\pm$2	&	8\\
	&111&	1.5$\pm$0.8	&	13\\		
	&408&	0.09$\pm$0.01	&	2	\\	
	&610&	0.32	&	2		\\
	&925&	0.04	&	2		\\
	&1420&	0.44	&	2		\\
	&1710&	0.26	&	2		\\
	\\
B1054$-$62&	410&	23$\pm$2.5&		1\\	
	&1382.5&	8.52$\pm$0.35&		2\\
	&1560	&8.46$\pm$0.24	&	2		\\ 
                       \hline 
\multicolumn{4}{c}{$^*$ full list of references at the beginning of the Appendix}
\end{tabular}
\end{table}

\begin{table}
\centering{\bf Table~\ref{tau_sc_table}.} Scatter time measurements (continued).
\vskip1mm
\begin{tabular}{lrlc}\hline
Pulsar & $f_{\rm obs}$[MHz] & $\tau_d$ [ms] & Reference$^*$ \\
\hline
\hline
B1114$-$41&	160	&7.4$\pm$2.9&		1\\
	&658&	1.36$\pm$0.62	&	2 \\	
	&660	&0.89$\pm$0.09	&	2		\\
	&1318.5&	0.08$\pm$0.01&		12 \\
	\\
B1154$-$62&	410&	27$\pm$1.7&		1\\	
	&685	&3.86$\pm$0.26		&12	\\
	&1327	&0.12$\pm$0.01		&12	\\	
	&1590	&0.07$\pm$0.01		&12\\
	\\
B1302-64	&410&	25$\pm$2.1&		1\\	
	&1382.5	&0.58$\pm$0.06		&2\\
	\\
B1323$-$58&	410 &	77$\pm$15&		1\\	
	&750&	30$\pm$0.27&		3\\	
	&1431&2.21$\pm$0.23&		12\\		
	&1560	&0.08$\pm$0.02&		2	\\	
	&1676&	1.38$\pm$0.38	&	12\\
	\\
B1323$-$62&	410&	185$\pm$35		&1\\
	&685&	19.42$\pm$0.15&		12\\	
	&750&	30$\pm$0.27		&3		\\
	&1318&	0.4$\pm$0.04		&12		\\
	&1347&	0.15		&12		\\
	&1382.5&	0.5	&	2		\\
	&1490.2	&0.58	&	2	\\	
	\\
B1356$-$60&	410	&24$\pm$1.7&		1	\\
	&660	&5.05$\pm$0.05&		12		\\
	&1374&	0.23		&	12		\\
	&1693&	0.57$\pm$0.04&		12		\\
	&2350&	0.01		&12\\
	\\
B1508+55&	44&	15$\pm$4&	13\\	
	&63&	6.4$\pm$4	&13\\
	&408&	0.62$\pm$0.24&	2\\		
	&610&	0.18$\pm$0.002&	2	\\	
	&	&0.14		&2		\\
	&	&0.35		&2		\\
	&925&	0.5	&	2\\		
	&1408&	0.63$\pm$0.002	&2	\\
	&1642&	0.35&	2\\		
	&4750&	5.26$\pm$1.9	&2\\
	\\
	B1557$-$50&	325&	411$\pm$150&		9\\	
	&610	&43$\pm$3		&9		\\
	&750	&16$\pm$0.25		&3	\\
	&1060&	5$\pm$0.6		&9	\\
	&1400&	1.98$\pm$0.35		&9\\		
	&1600&	0.97$\pm$0.24		&9\\
\hline
\multicolumn{4}{c}{$^*$ full list of references at the beginning of the Appendix}
	\end{tabular}
\end{table}

\begin{table}
\centering{\bf Table~\ref{tau_sc_table}.} Scatter time measurements (continued).
\vskip1mm\begin{tabular}{lrlc}\hline
Pulsar & $f_{\rm obs}$[MHz] & $\tau_d$ [ms] & Reference$^*$ \\
\hline
\hline
	B1558$-$50& 410&    20.8$\pm$2.2&  1\\
	 &  1374&  0.55 &           12\\
	  & 1518 & 1.88$\pm$0.78 & 12\\
         &1693 &  0.1& 12 \\
         \\
         B1620$-$42	&410	&36$\pm$4.7&		1\\	
	&1440	&3.17$\pm$1.93	&	2\\
	\\
B1641$-$45&	610&	73.4$\pm$4.8&		9	\\.
	&750&	40			&3		\\
	&1060&	8.8$\pm$0.4		&9\\		
	&4850&	2.5$\pm$0.4		&9\\
\\
B1642$-$03&	63	&7.2$\pm$1.2&		13	 \\
	&80	&4.4$\pm$0.11		&8\\
	&111	&1.5$\pm$0.5		&13	\\	
	&408	&0.9$\pm$0.02		&2\\
\\
J1705$-$3950&	610&	35$\pm$9& 14\\	
	&1374	&3.89$\pm$0.23&		12\\
\\
J1723$-$3659&	610	&17.2$\pm$2.5&	14\\	
	&1374&	1.3$\pm$0.32&		12	\\	
	&1518&	1.03$\pm$0.63	&	12\\	
\\
B1737$-$39	&410&	11$\pm$0.8&		1	\\
	&658&	4$\pm$0.09	&	2	\\	
	&660	&4.39$\pm$0.13&		12\\
\\
J1739$-$3023&	610&	0.28&		14\\	
	&1374&	0.03$\pm$1.37		&12	\\	
	&1518&	0.04$\pm$4.15		&12\\
\\
J1744$-$3130&	610	&11.8$\pm$6.0&	14\\
\\
B1745$-$12	&111&	55$\pm$24	&13\\
	&610&	5.06$\pm$0.2	&2\\		
	&1382.5&	7.23$\pm$0.56&	2\\
\\
B1749$-$28&	80	&24.9$\pm$2 &1\\
	& &40$\pm$5&		8	\\
	&160&	2.9$\pm$0.1&		1\\		
	&&	3.1$\pm$0.5&		8	\\	
	&408	&1.58$\pm$0.01&		2\\
	&610&	1.26$\pm$0.05&		2	\\
	&925&	1.83$\pm$0.058&		2\\	
	&1408&	2.02$\pm$0.035&		2\\		
	&1642	&0.31$\pm$0.09&		2		\\

								\hline
\multicolumn{4}{c}{$^*$ full list of references at the beginning of the Appendix}
\end{tabular}

\end{table}

\begin{table}
\centering{\bf Table~\ref{tau_sc_table}.} Scatter time measurements (continued).
\vskip1mm
\begin{tabular}{lrlc}\hline
Pulsar & $f_{\rm obs}$[MHz] & $\tau_d$ [ms] & Reference$^*$ \\
\hline
\hline
B1758$-$03&	111&	55$\pm$30&	13		\\
	&610	&0.12$\pm$1.8	&2\\
	B1758$-$23&	1274.75&	130.53$\pm$5.38&		12	\\
	&1374&	102.49$\pm$1.13	&	12\\	 	
	&1400&	111$\pm$19	&	10		\\
	&&	99$\pm$19	&	4		\\
	&1421&	83.22$\pm$3.92	&	12\\	 	
	&1518&	74.32$\pm$1.27&		12	\\	
	&1642&	55$\pm$10	&	10		\\
	&&	51$\pm$10	&	9		\\
	&2263&	17.87$\pm$0.84	&	12\\		
	&2600&	0.75$\pm$0.34	&	9	\\
	&4850&	0.23$\pm$0.08	&9\\
\\
B1802+03	&102	&33$\pm$10	&13\\
\\
B1818$-$04&	102&	60$\pm$40&		13\\
	&160&	50.3$\pm$5&		1\\		
	&     &	65$\pm$13	&	8		\\
	&410&	4.74$\pm$0.07&		2\\		
	&610&	2.82$\pm$0.03	&	2	\\	
	&925&	3.47$\pm$0.12	&	2		\\
	&1408&	3.28$\pm$0.09	&	2\\
	&1642&	3.47$\pm$0.16	&	2	\\
\\
B1821$-$19	&325&	57$\pm$23&		10\\
	&408&	22$\pm$4&		10\\		
	&410&	17$\pm$1.3&		1\\		
	&610&	4.6$\pm$0.2	&	10	\\			
	&925&	0.92$\pm$0.04&		2	\\		
	&1642&	0.00057$\pm$0.00018&		2\\
		B1831$-$04&	102	&50$\pm$25&	13\\
\\
B1839-04	&111&	150$\pm$15	&13\\
\\
J1841-0345&	610	&15.3$\pm$5.2&	14\\	
\\
B1844$-$04&	160&	63.5$\pm$30&		1\\
	&408&	15.4$\pm$0.53		&2	\\	
	&610&	0.77$\pm$0.08		&2		\\
	&925&	0.34$\pm$0.04		&2	\\	
	&1408&	0.04$\pm$0.01		&2\\		
\\
B1845$-$01&	235&	262$\pm$181	&	10	\\
	&408	&54$\pm$8		&10	 	\\
	&410	&42.6$\pm$3.4	&	2		\\
	&430&	42$\pm$6		&5		\\
	&610&	16$\pm$0.19	&	2		\\
	&&	17$\pm$1.4	&	10		\\
	&925&	5.71$\pm$0.36&		2\\		
	&1408&	5.95$\pm$0.32	&	2		\\
	&1642&	8.41$\pm$0.53	&	2\\		
	&4750&	0.16$\pm$0.05	&	2	\\	
		\hline
\multicolumn{4}{c}{$^*$ full list of references at the beginning of the Appendix}
\end{tabular}
\end{table}

\begin{table}
\centering{\bf Table~\ref{tau_sc_table}.} Scatter time measurements (continued).
\vskip1mm
\begin{tabular}{lrlc}\hline
Pulsar & $f_{\rm obs}$[MHz] & $\tau_d$ [ms] & Reference$^*$ \\
\hline
\hline
B1846-06&	111	&100$\pm$40&	13	\\
	&610&	3.42$\pm$0.5&	2		\\
	&	&3.42$\pm$0.45	&2		\\
	&	&3.39$\pm$0.43	&2		\\
	&657	&3.47$\pm$0.83&	2\\
\\
J1852-0635&	610&	15.1$\pm$3.6&	14\\	
	&1374	&0.95$\pm$0.0004&		12\\
\\
B1853+01	&111&	26$\pm$10	&13\\	
	&606& 	0.97$\pm$0.44&	2\\
\\
B1859+03	&320&	187	&		3\\
	&325&	241$\pm$51	&	10		\\
	&400&	83.93$\pm$2.07	&	2\\		
	&408&	89$\pm$9		&10		\\
	&410&	68.3$\pm$2.29	&	2	\\	
	&430&	65$\pm$11		&    5		\\
	&606&	12.98$\pm$0.18	&	2\\		
	&610&	12.6$\pm$0.9	&	10	\\	
	&925&	1.92$\pm$0.47	&	2		\\
	&1408&	0.13$\pm$0.01	&	2		\\
	&1420&	2$\pm$0.13		&2		\\
	&1642&	0.06$\pm$0.01		&2\\
\\
B1900+01&	243&	176$\pm$58&		10	\\ 
	&325&	34$\pm$2		&10		\\
	&408&	14.7$\pm$1.3&		10	\\	
	&410&	13.04$\pm$0.39&		2\\		
	&430&	11$\pm$2		&5		\\
	&610&	3.37$\pm$0.07&		2	\\
	&925&	1.02$\pm$0.12	&	2	\\
	&1408&	0.71$\pm$0.07	&	2\\
\\
B1907+02&	111&	140$\pm$50	&	13\\	
	&160&	27.9$\pm$16&		1		\\
	&320&	3$\pm$1		&6		\\
	&410&	2.84$\pm$0.79&		2\\
	&610&	3.66$\pm$0.34&		2	\\	
	&1408&	5.76$\pm$0.68&		2\\
	\\
B1907+10&	160&	26.5$\pm$8.1&		1\\
	&&	27$\pm$7		&8		\\
	&320&	2.2$\pm$0.7&		6		\\
	&410&	0.13$\pm$0.05&		2	\\
\\
B1911$-$04	&102&	35$\pm$15		&13\\	
	&160&	16.7$\pm$1.8		&1	\\	
	&&	32$\pm$5		&8		\\
	&408&	0.861$\pm$0.6		&2	\\	
	&610&	0.72$\pm$0.81		&2	\\	
	&925&	0.47$\pm$0.96		&2	\\	
	&1408&	0.12$\pm$0.004		&2	\\
	&1642&	0.26$\pm$1.96		&2	\\	
	&4850&	0.09		&2\\
		\hline
		\multicolumn{4}{c}{$^*$ full list of references at the beginning of the Appendix}
\end{tabular}
\end{table}

\begin{table}
\centering{\bf Table~\ref{tau_sc_table}.} Scatter time measurements (continued).
\vskip1mm
\begin{tabular}{lrlc}\hline
Pulsar & $f_{\rm obs}$[MHz] & $\tau_d$ [ms] & Reference$^*$ \\
\hline
\hline
B1913+10&	408&	23.87$\pm$0.77&		2\\	
	&430&	14.6			&7		\\
	&610&	5.41$\pm$0.15		&2		\\
	&925&	1.12$\pm$0.37		&2		\\
	&1408&	1.56$\pm$0.011		&2		\\
	&1420&	2.61$\pm$0.29		&2		\\
	&1642&	0.086$\pm$0.052		&2\\
\\
B1914+13&	320&	5.5$\pm$2.5&		6\\
	&610&	1.18$\pm$0.27		&2		\\
	&1408&	0.39$\pm$0.01		&2\\
\\
B1915+13&	102&	40$\pm$20&		13\\	
	&160&	11.7$\pm$1.9		&1		\\
		&	&12$\pm$3		&8		\\
	&408&	1.66$\pm$0.07&		2\\		
	&925&	1.76$\pm$0.07	&	2	\\	
	&1642&	1.22$\pm$0.15	&	2\\
\\
B1919+20	&430	&11.7&			5\\
	&606	&0.13	&2 \\	
\\
B1919+21&	44&	14.9$\pm$3.6&	13\\
	&63&	4.6$\pm$3.2&	13\\	
	&111&	0.6$\pm$0.4	&13\\		
	&408&	0.33$\pm$0.002&	2\\		
	&610&	0.35$\pm$0.001	&2\\
\\
B1920+21&	102.75&	26&			2	\\
	&160	&96.8$\pm$50	&	1		 \\
	&243	&4.4$\pm$1.5&		10		\\
	&410	&5.11$\pm$0.95&		2	\\	
	&925	&8.63$\pm$1.37	&	2\\
\\
B1929+20&	410	&10.58$\pm$0.54&		2\\
	&430	&4.83			&7		\\
	&925	&0.007		&2		\\
	&1408&	0.04$\pm$0.04&		2\\
\\
B1933+15	&430&	5.8			&7\\
\\
B1933+16&	110	&67&			3	\\
	&111&	50$\pm$15	&	13		\\
 	&160&	21.7$\pm$1.6&		1	\\	
	&&	25$\pm$4&		8		\\
	&243&	4.6$\pm$0.2&		10		\\
	&325&	1.8$\pm$0.1&		10		\\
	&410&	1.62$\pm$0.22&		2		\\
	&610&	0.51$\pm$0.33	&	2		\\
	&925&	0.05$\pm$0.01	&	2	\\	
	&1408&	0.02$\pm$0.01	&	2\\
		\hline
		\multicolumn{4}{c}{$^*$ full list of references at the beginning of the Appendix}
\end{tabular}
\end{table}

\begin{table}
\centering{\bf Table~\ref{tau_sc_table}.} Scatter time measurements (continued).
\vskip1mm
\begin{tabular}{lrlc}\hline
Pulsar & $f_{\rm obs}$[MHz] & $\tau_d$ [ms] & Reference$^*$ \\
\hline
\hline
B1940$-$12&160&	4.4$\pm$1.1	&	1		\\\
	&410&	0.93	&2		\\
	&610&	1.05$\pm$0.07		&2		\\
	&1318&	1.66$\pm$3.58		&12		\\
	&1642&	2.26$\pm$2.69		&2\\
	B1943+18	&430	&15$\pm$5		&6\\
\\
B1946+35&	111&	45$\pm$10	&	13	\\
	&320&	32	&		3		\\
	&408&	16$\pm$0.12&		2		\\
	&430&	15$\pm$3		&5		\\
	&610&	3.71$\pm$0.1&		2\\		
	&925&	0.6$\pm$0.57&		2	\\	
	&1642&	0.4$\pm$0.06	&	2\\
\\
B1953+50 &	111&	45$\pm$10&	13\\		
	&408&	1.55$\pm$0.25&	2	\\
	&925&	0.18$\pm$0.09  &2\\
\\
B2002+31&	243&	27$\pm$13&		10\\
	&320&	10			&3		\\
	&325&	5.3$\pm$1.2&		10		\\
	&408&	2$\pm$1.4	&	10		\\ 
	&	&5.51$\pm$0.76		&2		\\
	&610&	2.62$\pm$1.09&		2		\\
	&925&	0.033$\pm$0.049&		2\\		
	&1408&	5.33$\pm$0.41	&	2	\\			
	&1642&	4.13$\pm$1.08	&	2\\
\\
B2020+28&	63	&2.3$\pm$1&	13	\\	
	&111&	0.5$\pm$0.3	&13	\\	
	&410&	0.83$\pm$0.08	&2\\
\\
B2027+37&	111&	132$\pm$10&	13\\
	&610&	3.94$\pm$0.86&	2\\		
	&&	0.5	&2\\		
	&&	3.85$\pm$0.81	&2\\		
	&1408&	7.15$\pm$0.72&	2	\\	
	&&	7.16$\pm$0.74&	2	\\	
	&1420&	6.02$\pm$0.48&	2\\
\\
B2053+36&	111&	74$\pm$25	&	13\\
	&320	&7.7$\pm$1.5	&	6\\
	&430	&2.2$\pm$0.4	&6\\
	&925	&0.13$\pm$0.01&		2\\
\\
B2111+46 &	111&	120$\pm$40	&13\\		
	&408&	1.5$\pm$2.71	&2\\		
	&610&	5.19$\pm$0.71&	2	\\	
	&&	2.34$\pm$0.23	&2		\\
	&1408	&2.67$\pm$0.39	&2\\
		\hline
		\multicolumn{4}{c}{$^*$ full list of references at the beginning of the Appendix}
\end{tabular}
\end{table}

\begin{table}
\centering{\bf Table~\ref{tau_sc_table}.} Scatter time measurements (continued).
\vskip1mm
\begin{tabular}{lrlc}\hline
Pulsar & $f_{\rm obs}$[MHz] & $\tau_d$ [ms] & Reference$^*$ \\
\hline
\hline
B2113+14&	102&	25$\pm$10	&13	\\
	&1408&	0.11$\pm$0.03&	2\\
	\\
B2217+47	&44&	70$\pm$20&	13\\
	&63	&12$\pm$9	&13	\\
	&111&	3.2$\pm$0.3	&13\\		
	&410&	0.42$\pm$0.23	&2	\\
	&&		0.37$\pm$1.94	&2		\\
	&610	&0.47$\pm$4.69	&2		\\
	&1408&	0.5	&2\\
\\
B2303+46	&325&	1.06$\pm$0.0005	& 14\\
	&610	&0.65$\pm$0.03		&	14\\
	&1060&	0.35& 14 \\
\\
B2303+30&	44&	300$\pm$100&		13\\
	&63&	110$\pm$20		&13		\\
	&111&	13$\pm$3		&13		\\
	&160&	9.9$\pm$3.6	&		1	\\
	&410&	1.47	&		2	\\	
	&610&	1.66		&	2	\\	
	&1408&	0.631		&	2\\
\\
B2351+61&	111&	35$\pm$5&	13 \\	
		\hline
		\multicolumn{4}{c}{$^*$ full list of references at the beginning of the Appendix}
\end{tabular}
\end{table}
\end{document}